\documentclass[psfig,preprint]{aastex}
\usepackage{emulateapj5}

\newif\ifAMStwofonts
\AMStwofontstrue

\def\kms{km~s$^{-1}$}

\def\ga{\mathrel{\hbox{\rlap{\hbox{\lower4pt\hbox{$\sim$}}}\hbox{$>$}}}}
\def\la{\mathrel{\hbox{\rlap{\hbox{\lower4pt\hbox{$\sim$}}}\hbox{$<$}}}}

\shorttitle{Compact HI clouds from the GALFA-HI survey}

\shortauthors{Begum et al.}
\begin{document}

\title{Compact HI Clouds from the GALFA-HI Survey}

\author{Ayesha Begum$^1$,  Sne\v{z}ana Stanimirovi\'{c}$^1$, Joshua E.
Peek$^2$, Nicholas, P. Ballering$^1$,
Carl Heiles$^3$, Kevin A. Douglas$^4$,
Mary Putman$^2$, Steven J. Gibson$^5$, Jana Grcevich$^2$, Eric J.
Korpela$^6$, Min-Young Lee$^1$, Destry Saul$^2$, John S. Gallagher$^1$ III}
\affil{
$^1$University of Wisconsin, Madison, 475 N Charter St, Madison, WI 53703,
$^2$Department of Astronomy, Columbia University, New York, NY 10027, USA,
$^3$Radio Astronomy Lab, UC Berkeley, 601 Campbell Hall, Berkeley, CA 94720
$^4$ School of Physics, University of Exeter, Stocker Road, Exeter,
United Kingdom EX4 4QL
$^5$ Department of Physics and Astronomy, Western Kentucky University,
Bowling Green, KY 42101
$^6$ Space Sciences Laboratory, University of California, Berkeley,
CA 94720
}
\email{sstanimi@astro.wisc.edu, heiles@astro.berkeley.edu}

\begin{abstract}
The Galactic Arecibo L-band Feed Array HI (GALFA-HI) survey
is mapping the entire Arecibo sky at 21-cm, over a
 velocity range of $-$700 to +700 km s$^{-1}$ (LSR), at a
  velocity resolution of 0.18 km s$^{-1}$ and a spatial
resolution of 3.5 arcmin.
The unprecedented resolution and sensitivity of the GALFA-HI survey
have resulted in
the detection of  numerous isolated, very compact HI clouds at low Galactic
velocities, which are distinctly separated from the HI disk emission.
In the limited area
of $\sim$4600 deg$^2$ surveyed so far, we have detected 96 of such compact clouds. The
detected clouds are  cold with a median
T$_{k,max}$ (the kinetic temperature in the case in which
there is no non-thermal
broadening) of 300 K. Moreover, these clouds  are quite compact and faint,
with median values of 5 arcmin in angular size, 0.75 K in peak brightness
temperature, and  $5\times10^{18}$ cm$^{-2}$ in HI column density. Most of the
clouds deviate  from Galactic rotation at the 20-30 km s$^{-1}$ level, and a
significant fraction show evidence for a multiphase medium and velocity
gradients. No counterparts for these clouds were found in other wavebands.
From  the modeling of spatial and velocity distributions of the whole compact
cloud population, we find that the bulk of the compact clouds are related
to the Galactic disk, and their distances are likely to be in the
range of 0.1 to a few kpc. We discuss various possible scenarios for 
the formation and maintenance of this cloud population and its 
significance for Galactic ISM studies.
 \end{abstract}

\keywords{ISM: clouds --- ISM: structure --- radio lines: ISM}

%\tableofcontents

\section{Introduction}

%The compact clouds may thus represent
%the low-column density examples of the population of CNM Galactic 
%clouds on sub-pc scale.

The neutral interstellar medium (ISM) is known to exist in 
two thermal equilibrium states: the cold neutral medium (CNM) and 
the warm neutral medium (WNM)\citep{Field69,McKee77a,Wolfire03}. 
Both the CNM and the WNM are
%\cite{McKee77a} 
%summarized typical properties of the CNM clouds: a kinetic temperature
%$\sim$ 80 K, a clouds size $\sim$ 2 pc and 
%an HI column density $\sim$ 10$^{20}$ cm$^{-2}$. 
highly structured over a range of spatial scales, 
and exhibit a variety of morphologies including sheets, filaments, shells
and clouds \citep{kalberla09}.
The 21-cm line of neutral hydrogen (HI) has been the main tracer
of the multiphase neutral ISM in the Galactic thin disk and
beyond. 
The disk-halo interface region, extending up to 1--1.5 kpc above the disk
and interfacing with the Galactic halo gas, was initially considered to 
represent a smooth envelope of HI surrounding
the Galactic spiral structure.
However, with the improvement in sensitivity and spatial resolution of the
Galactic surveys, discrete HI clouds have been detected down to 
a scale of a few parsecs 
\citep{lockman02,stil06b,snez06,ford08}.

Traditionally, HI observations have been able to trace 
the entire hierarchy of structures in the neutral ISM  on 
scales $\geq$ 1 pc. However, the small-scale end 
of this spectrum, i.e scales $<$ pc, is still largely unexplored because 
of a paucity of high spatial/velocity resolution imaging surveys. 
Cold HI structures have been observed on sub-pc scales, predominantly 
within the Galactic plane, as HI self-absorption (HISA) features against 
the extended  background of the WNM \citep{gibson05}. On the other hand, 
detection of the smallest HI structures down to tens of astronomical units 
mainly rely on the absorption measurements against continuum background 
sources, or from  the time variability of HI absorption profiles against 
pulsars \citep{frail94,Heiles97,snez07}.
Regarding the sub-pc scale structure of the Galactic disk-halo interface region,
several studies have been conducted by mapping the HI gas in emission 
using interferometers. However, such studies require a substantial investment 
of telescope time, hence have been mainly limited to small, 
targeted regions in the Galaxy \citep{stil06a,dedes08,bekhti09,dedes10}.

The presence of sub-pc scale clouds in the ISM raises many important questions.
For example, how abundant is this cloud population? 
What are the formation and survival mechanisms for such clouds? 
Also, what role these clouds play in the general ISM?  
In addition, such clouds are fascinating as they may be related to major 
dynamical processes in the ISM such as stellar winds \citep{matthews08,gerard06},
shocks \citep{gibson05}, turbulence \citep{Semadeni06,avillez05,Audit05} and
Galactic accretion \citep{heitsch09}. To fully investigate the nature
of small-scale HI structure a sensitive, unbiased, high resolution survey of the
entire sky is required.

%Further, such compact HI clouds could also be potential ultra-faint dwarf
%galaxies, like the recently discovered gas-rich dwarf Leo-T \citep{leot}.

The Galactic Arecibo L-band Feed Array HI (GALFA-HI) survey is
successfully mapping the entire Arecibo sky at 21 cm. The survey
covers a velocity range of $-$700 to +700 km s$^{-1}$ (LSR)
at an unprecedented velocity resolution of 0.18 km s$^{-1}$ and
a spatial resolution of 3.5 arcmin. The combination of sensitivity
and resolution provided by the GALFA-HI survey allows us to probe a new
regime of faint, small HI objects that have not been seen before
in lower resolution surveys e.g. Leiden/Argentine/Bonn (LAB) survey and Galactic
All Sky Survey (GASS) \citep{lab,gass} or lower sensitivity surveys e.g. 
Canadian Galactic Plane Survey (CGPS) \citep{stil06a}.
In particular, in this paper we focus on a population of strikingly 
compact and isolated HI clouds detected primarily  at Galactic velocities.
%In this paper we present the properties of compact clouds 
%detected at Galactic velocities in the GALFA-HI survey, and discuss their
%origin and their role in Galactic ISM studies.

The structure of this paper is organized in the following way.
Section~\ref{s:observations} summarizes the GALFA-HI observations and
data reduction.
In Section~\ref{sec:cloud-catalog} we describe our search for compact HI clouds
and provide a catalog of basic cloud properties.
The cloud properties, their distribution models  and distance
constraints are described in
Sections~\ref{sec:properties}$-$\ref{sec:tangent}, whereas
Section~\ref{s:discussion}
discusses various possible origins of this cloud population.

\section{Observations and Data Processing}
\label{s:observations}

The GALFA-HI survey consists  of many individual projects, being
observed in both solo modes and commensally with ALFALFA (Arecibo
Legacy Fast ALFA Survey; \cite{alfalfa}), AGES (Arecibo Galaxy
Environment Survey; \cite{ages}) and GALFACTS (Galactic ALFA Continuum
Transit Survey; \cite{galfacts}) as the TOGS (Turn-On GALFA
Spectrometer) and TOGS2 projects. Data cubes used for this study have
been made by the core GALFA-HI
group\footnote{http://sites.google.com/site/galfahi/Home}
and are based on the first data release (released separately as
``Spring" and ``Fall"
data cubes), which uses data from TOGS and several smaller, targeted
GALFA-HI projects. The Spring data cubes  contain 1030 hours of
Arecibo observing time and cover over $\sim$2550 deg$^2$ including the
projects TOGS and a2220 (PI: J. Peek). The Fall data cubes contain
830 hours over $\sim$2050 deg$^2$ comprising the projects TOGS, a2050 (PI:
Peek \& Heiles) and a2172 (PI. C. Heiles). The observing method used
is drift scanning for TOGS and basketweave scanning for the other
projects. The details of the observing modes are summarized in
\cite{snez06} and Peek \& Heiles (2008).

To generate the HI data cubes, raw data obtained with the Arecibo
L-band Feed Array (ALFA) were reduced using the GALFA-HI standard
reduction pipeline, the details of which are described in Peek \&
Heiles (2008). The typical RMS noise is in the range of  75$-$120 
mK (per 0.736 \kms~velocity channels), depending on the
area covered under different GALFA-HI projects. 
The data used in this study have not been corrected for the
first sidelobe. From our analysis, we find that this effect is small
and does not have a significant impact on our results. For example,
the first-sidelobe correction leads to an increase in the peak
brightness temperature of $<10-15$\%, a decrease in the velocity
linewidth by $\sim$5\%, and a decrease in the cloud size by
$\sim$ 10-15\% .

\begin{figure*}
%\epsscale{2.0}
%\plottwo{Coverage_fall1.ps}{Coverage_fall2.ps}
\includegraphics[scale=0.7,angle=-90.0]{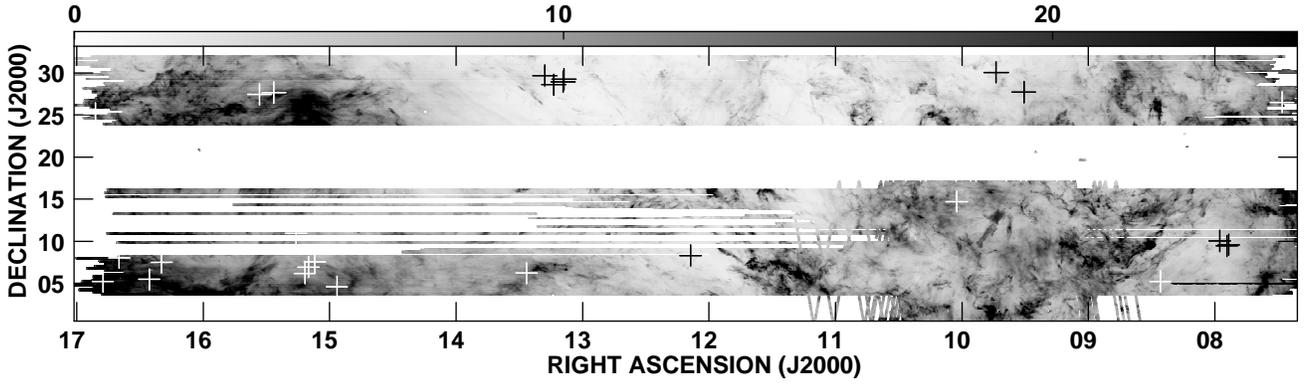}
\caption{\label{f:spring} Peak brightness temperature image made using
GALFA-HI data for the "Spring" portion of the survey. 
The velocity range used for making the map is $-120
<{\rm{V_{LSR}}}<$120 km s$^{-1}$. The greyscale units are Kelvin.
The crosses show the location of detected compact clouds. The size of
the crosses has been scaled by a factor of $\sim$30 for visual
clarity. The color of the crosses is chosen to get a good 
contrast with the background greyscale.
}
\end{figure*}

\begin{figure*}
%\epsscale{2.0}
%\plottwo{Coverage_fall1.ps}{Coverage_fall1.ps}
\includegraphics[scale=1.00]{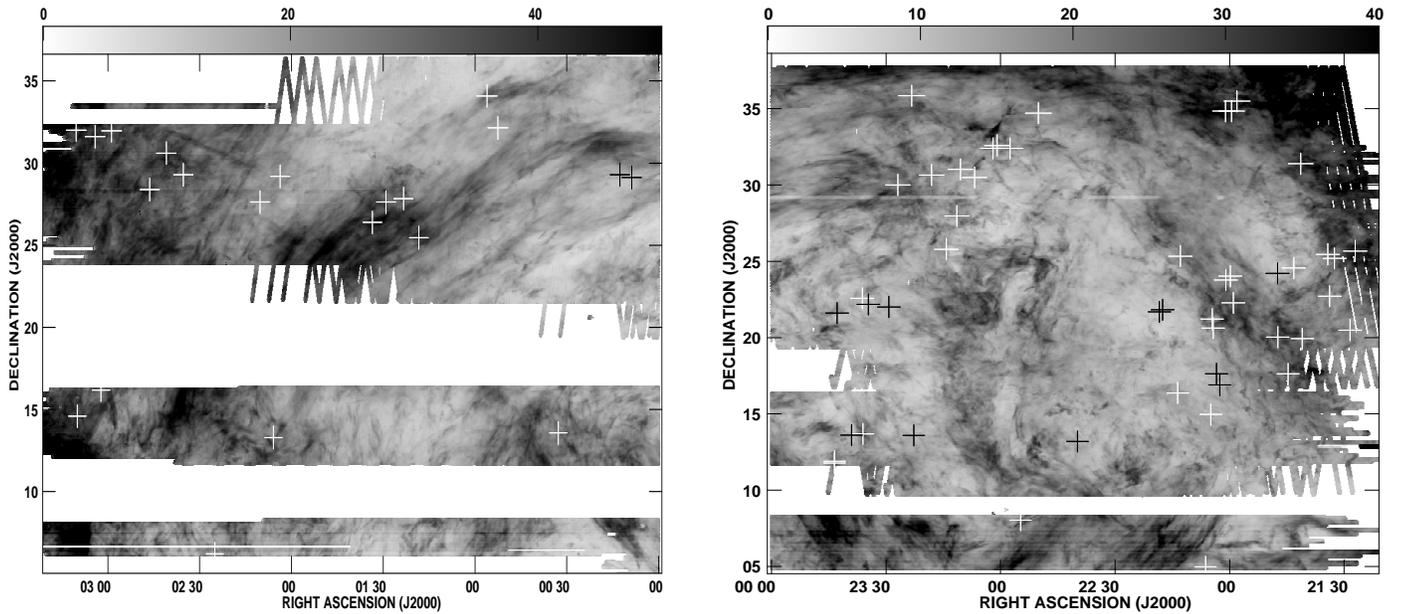}
\caption{\label{f:fall} Peak brightness temperature map made using
GALFA-HI data for the "Fall" portion of the survey. 
The velocity range used for making the map is $-120
<{\rm{V_{LSR}}}<$120 km s$^{-1}$. The greyscale units are Kelvin.
The crosses show the location of detected compact clouds. The size of
the crosses has been scaled by a factor of $\sim$16
for visual clarity. The color of the crosses is chosen to get a good 
contrast with the background greyscale. 
}
\end{figure*}

Figures~\ref{f:spring} and \ref{f:fall} show the sky coverage of the
GALFA-HI data used in this study for the ``Spring" and ``Fall" portion 
of the survey.  There are still significant gaps in
the survey coverage, resulting in a non-uniform sampling of the RA-Dec
space.
In terms of Galactic coordinates, the current survey coverage consists
of three main areas: $l=0-60$ degrees and $b \sim 30$ to $60$ degrees, 
$l=60-180$ degrees and $b \sim -50$ to $-20$ degrees and $l=200-250$ 
degrees and $b \sim 20$ to $50$ degrees.
The total angular search area used for our compact cloud study is $\sim$4600
deg$^2$, though on the completion of the GALFA-HI survey an area of
~13000 deg$^2$ will be available for an expanded search.

\section{Compact HI Clouds}
\label{sec:cloud-catalog}

\subsection{Search method}
\label{sec:search}

The standard survey data cubes were combined to produce two sets of
larger cubes which are easier for visual inspection,  viz.~(i) low
velocity resolution data cubes of size 40 degrees $\times$ 18 degrees
$\times$ 1500 \kms~with a velocity resolution of
7.4 \kms,  and (ii) high velocity resolution data
cubes of 40 degrees $\times$ 18 degrees $\times$ 240 \kms~ with a
velocity resolution of 1.8 \kms.
Each three-dimensional (RA-Dec-Velocity) datacube was visually
inspected using the visualization program KVIS (part of KARMA, Gooch
1996). The data cubes were searched for {\it compact} and {\it
isolated} clouds. The  clouds which appeared to be a part of some
larger,  filamentary structure were not considered.

The first round of the visual search was performed using the low
velocity resolution data cubes over a wide velocity range of
$-750.0~<~{\rm{V_{LSR}}}~<750.0$ km s$^{-1}$. The compact clouds
of interest were mainly found within the velocity range of
$-120.0~<~{\rm{V_{LSR}}}~<120.0$ \kms.
Compact clouds found outside this range were relatively rare and were
generally identified as known galaxies. Therefore, in this paper we focus only on the
clouds with $-120.0~<~{\rm{V_{LSR}}}~<120.0$ \kms.

The second round of the search was performed using the high resolution
data cubes, restricting the search velocity range to
$-120.0~<~{\rm{V_{LSR}}}~<120.0$ \kms.
Each selected cloud was then  inspected in the original high velocity
resolution (0.18 \kms) data cube to reject any spurious or very faint
signals and also to measure individual cloud properties. In total, 96
clouds were identified. The cloud properties are presented and
discussed in the following sections.

\subsection{Cloud Catalog}
\label{sec:catalog}

We have compiled a catalog of 96 compact HI clouds in Table~\ref{t:table3}.
Column (1) is the cloud catalog number; (2) and (3) are Right
Ascension (RA) and Declination (Dec) in J2000; (4) and (5) are the Galactic
longitude ($l$) and latitude ($b$); (6) gives the size in
arcminutes, estimated as the geometrical mean of
the measured angular extent along
the major and minor axes ($\theta=\sqrt{\theta_{max} \times
\theta_{min}}$); 
(7) and (8) give the central LSR velocity (V$_{\rm LSR}$, measured in
\kms) of the fitted Gaussian components (only
one value is given if the velocity profile can be well fit with a
single Gaussian function);
(9) and (10) give the velocity FWHM linewidth(s) of the fitted Gaussian
components, $\Delta V$, in km~s$^{-1}$,
(11) and (12) are the peak brightness temperature(s) of the fitted
Gaussian components, T$^{pk}_b$, in Kelvin; and
(13) is the HI column density, N${\rm{_{HI}}}$, measured in units of
10$^{18}$ atoms cm$^{-2}$. N${\rm{_{HI}}}$ is derived under the
assumption that the gas is optically thin.

Cloud size  was determined using the total integrated HI column
density maps, by fitting an ellipse to the half-peak N${\rm{_{HI}}}$
contour, after background subtraction. No correction was applied to
the angular size for the convolution with the Arecibo main beam and
sidelobes. The integrated velocity profile for each cloud was obtained
by averaging over the area of the
cloud. For each profile, the background level was determined by
fitting a polynomial function
(typically a second, or third order, depending on the baseline shape)
to the line-free channels around the integrated line profile of the
cloud.  The baseline-subtracted velocity profile for each cloud was
then fitted with one or two Gaussian functions.

\begin{figure*}
%\epsscale{1.0}
%\plotone{l_V_lsr.ps}
\begin{center}
\includegraphics[scale=0.6,angle=-90]{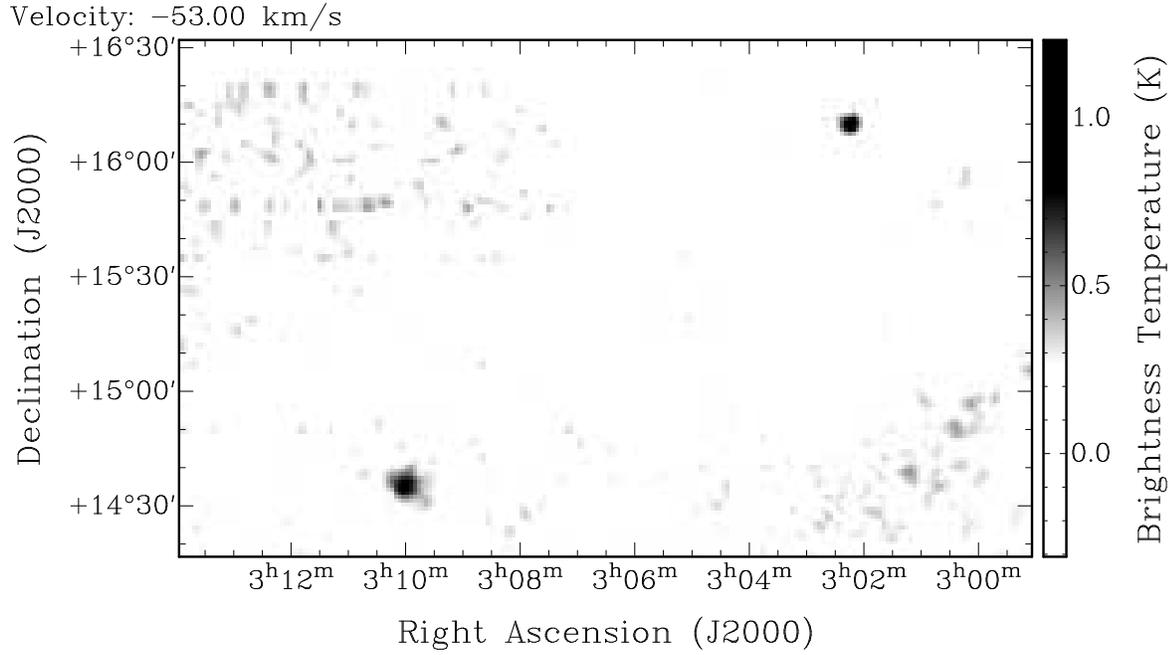}
\caption{\label{f:image1} A GALFA-HI image at V$_{\rm LSR}=-53$ \kms,
showing two example compact HI clouds, No 18 (upper right) and 20
(lower left). %The width of the velocity channel is 1.84 kms$^{-1}$.
}
\end{center}
\end{figure*}

\begin{figure*}
\epsscale{2.4}
\plotone{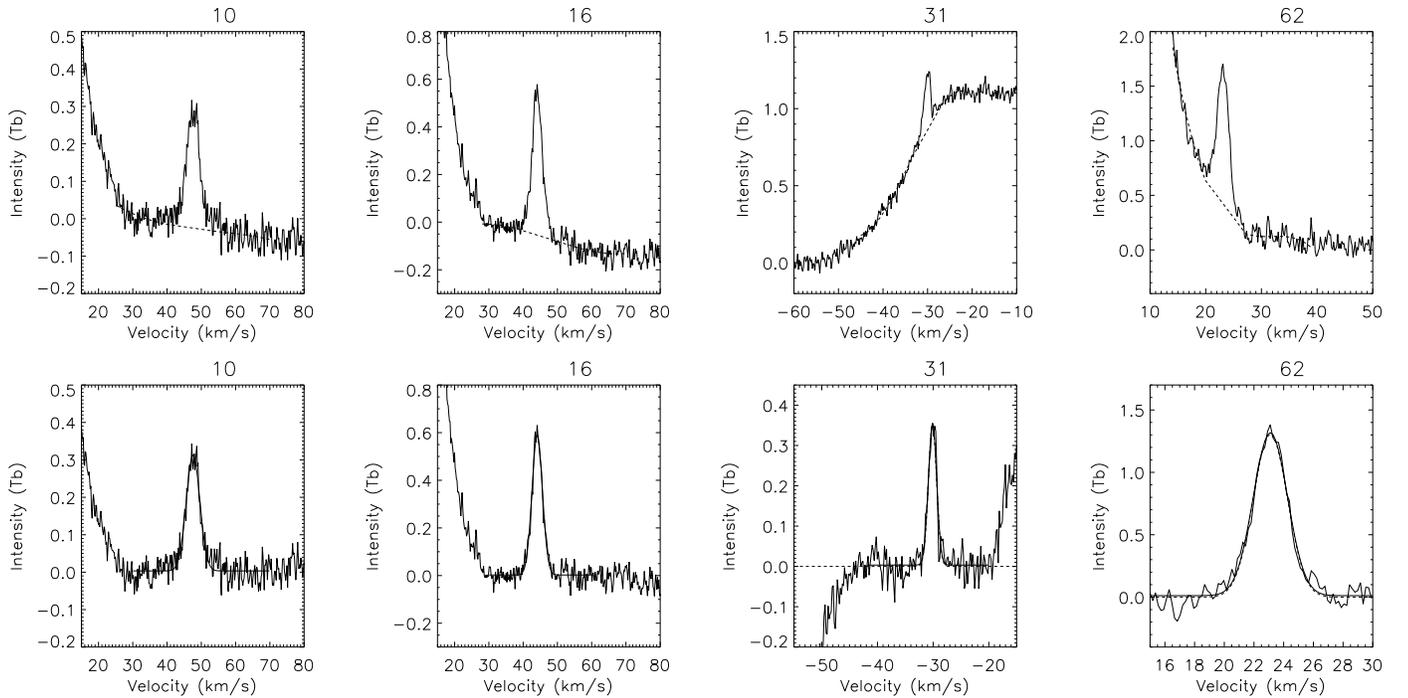}
\caption{\label{f:profile_fit} Four examples of HI velocity profiles
of compact HI clouds. The dashed line in the top panels shows the
estimated baseline, while the solid line in the bottom panels
represents Gaussian functions fitted to the HI line profile. }
\end{figure*}

The positions of all clouds, overlayed on the peak brightness
temperature image for both the Spring and Fall portions of the survey,
are shown in Figures~\ref{f:spring} and \ref{f:fall}. An example peak
brightness temperature image at a velocity of $-53$ \kms, showing two compact HI
clouds, is presented in Figure~\ref{f:image1}. 
Several examples of HI velocity profiles at the cloud center, with
estimated baselines and fitted Gaussian functions, are shown in
Figure~\ref{f:profile_fit}.

\begin{table*}
%\begin{tiny}
\begin{center}
\caption{The compact cloud sample with derived Gaussian parameters}
\label{table3}
\begin{tabular}{c c c c c c c c c c c c c}
\hline
\hline
    \multicolumn{1}{c}{\small Cloud} &
    \multicolumn{1}{c}{ RA } &
    \multicolumn{1}{c}{ Dec } &
    \multicolumn{1}{c}{ l } &
    \multicolumn{1}{c}{ b } &
    \multicolumn{1}{c}{ $\theta$ } &
    \multicolumn{1}{c}{v$^1_{LSR}$} &
    \multicolumn{1}{c}{v$^2_{LSR}$} &
    \multicolumn{1}{c}{$\Delta V^1$} &
    \multicolumn{1}{c}{$\Delta V^2$} &
    \multicolumn{1}{c}{ T$_b^{Pk_1}$ } &
    \multicolumn{1}{c}{ T$_b^{Pk_2}$ } &
    \multicolumn{1}{c}{ N$\rm{_{HI}}$ } \\
%  No& (h:m:s) & (d:m:s) &  ($^\prime$)& &(\kms)& (\kms)&(\kms) &
%(\kms)&(K) &(K)  & (10$^{18}$ cm$^{-2}$)\\
{\small   No}& {\small (J2000)} & {\small (J2000)} & (deg) &  (deg) &($^\prime$) &{\small (kms$^{-1}$)}& {\small (kms$^{-1}$)}&{\small (kms$^{-1}$)} & {\small ( kms$^{-1}$ )}&(K) &(K)  & {\small{(10$^{18}$}}   \\
   & (h:m:s) & (d:m:s) & & & && & & & &  & {\small cm$^{-2}$)} \\
\hline
\hline
1 & 00:08:46 & 29:07:30 & 111.83 & $-32.82$ & 4.97   & 25.84 & - & 4.20 & - & 0.19  & -
  & 1.19  \\
2 & 00:12:38 & 29:17:30 & 112.86 & $-32.82$ &7.12 &  $-38.41$ & - & 3.66 & - & 0.61  &
-   & 3.97 \\
3 & 00:32:46 & 13:34:30 & 116.00 & $-49.04$ & 3.74 &  47.21 & - & 17.59 & - & 0.32  &
-   & 8.26 \\
4 & 00:52:30 & 32:08:30 & 123.19 & $-30.72$ &6.93 &  $-49.98$ & $-50.53$ & 0.42 & 2.97 &
0.08  & 0.16   & 1.00  \\ 
5 & 00:56:06 & 34:05:30 & 124.03 & $-28.77$ &3.5 &  50.09 & - & 4.84 & - & 0.79  & -   & 6.30 \\
6 & 01:18:21 & 25:27:30 & 130.54 & $-37.01$ &5.12 &  $-50.60$ & $-47.50$ & 1.98 & 17.79 &
0.28  & 0.29   & 11.29 \\ 
7 & 01:23:22 & 27:50:30 & 131.50 & $-34.49$ &9.80 &  31.30 & - & 5.39 & - & 0.69  & -   & 6.92 \\
8 & 01:29:06 & 27:37:30 & 133.06 & $-34.50$ &5.61 &  $-56.14$ & - & 5.62 & - & 0.63  &
-   & 5.22 \\
9 & 01:33:34 & 26:23:30 & 134.53 &  $-35.52$ &3.50 &  28.79 & - & 4.46 & - & 0.39  & -
  & 2.59 \\
10 & 02:03:42 & 29:11:30 & 141.35 & $-32.08$ &5.59 &  47.60 & - & 4.64 & - & 0.34  &
-   & 2.79 \\
11 & 02:05:54 & 13:17:30 & 149.35 &  $-45.72$ &4.47 &  27.76 & - & 2.71 & - & 0.28  &
-   & 1.18 \\
12 & 02:10:14 & 27:37:30 & 143.56 &  $-32.06$ & 3.50  & $-52.08$ & $-51.95$ & 3.98 & 13.20
& 0.36  & 0.25   & 9.24 \\ 
13 & 02:25:02 & 06:12:30 & 160.55 & $-49.70$  & 5.59 &  $-43.90$ & - & 12.02 & - & 0.39  & -   & 6.31 \\
14 & 02:35:21 & 29:17:30 & 148.65 & $-28.33$   & 6.75 &  $-38.79$ & - & 3.29 & - & 0.65  &
-   & 3.69 \\
15 & 02:40:54 & 30:36:30 & 149.19 & $-26.62$  & 3.50 &  57.57 & - & 7.54 & - & 0.37  &
-   & 3.78 \\
16 & 02:46:22 & 28:23:30 &  151.55 & $-28.01$  & 5.48 &  44.11 & - & 3.60 & - & 0.63  &
-   & 4.11 \\
17 & 02:58:46 & 31:58:30 & 152.15 & $-23.57$  &4.0 &  $-60.02$ & - & 2.00 & - & 0.60  &
-   & 2.11 \\
18 & 03:02:14 & 16:10:30 & 163.01 & $-36.31$  &3.50   & $-51.91$ & $-50.38$ & 3.37 & 11.87
& 0.39  & 0.25   & 8.32 \\ 
19 & 03:04:10 & 31:36:30 & 153.42 & $-23.28$ &5.86 &  $-53.07$ & $-52.38$ & 1.86 & 5.51 & 1.74  & 0.19   & 8.21 \\ 
20 & 03:10:02 & 14:35:30 & 166.06 & $-36.41$ &3.50 &  $-53.46$ & - & 10.27 & - & 0.34  & -   & 4.87 \\
21 & 03:10:22 & 32:00:30 & 154.39  & $-22.24$  & 9.95 & $-48.07$ & $-51.98$ & 6.36 & 16.50
& 0.19  & 0.36   & 13.99 \\ 
22 & 07:28:10 & 26:30:30 & 192.32  &  19.30 & 6.71 & $ -84.36$ & $-80.11$ & 4.23 & 18.65 & 0.34  & 0.22   & 10.62 \\ 
23 & 07:53:30 & 29:34:30 &  211.25& 18.09  &5.48 &  $-31.81$ & - & 6.87 & - & 0.19  & - & 1.72 \\
24 & 07:54:22 & 09:26:30 & 211.48 &  18.23  &3.60   & $-29.68$ & - & 6.12 & - & 0.15  &
-   & 1.22 \\
25 & 07:57:37 & 10:02:30 &  211.27&   19.21 &3.74  & $-31.00$ & - & 0.85 & - & 0.48  &
-   & 0.77 \\
26 & 08:25:54 & 05:13:30 &  219.23 & 23.36   & 7.35 &  55.36 & - & 4.81 & - & 0.11  &
-   & 0.81 \\
27 & 09:30:38 & 27:42:30 &  200.06   & 45.75   & 3.97 &  $-120.64$ & - & 3.71 & - & 0.21
& -   & 1.33 \\
28 & 09:43:50 & 30:02:30 & 197.38 &  48.99 & 3.86 &  $-111.14$ & - & 1.06 & - & 0.15
& -   & 0.23 \\
29 & 10:02:29 & 14:43:20 & 221.70  & 48.86   & 6.12 &  $-43.76$ & - & 4.22 & - & 1.08  &
-   & 7.77 \\
30 & 12:08:46 & 08:18:30 &  272.73  &  68.64   &  4.62 &  61.76 & - & 13.26 & - & 0.59  &
-   & 9.78 \\
31 & 13:08:46 & 29:14:30 &  62.90  & 85.63   & 5.74 &  $-30.11$ & - & 1.75 & - & 0.35  &
-   & 1.19 \\
32 & 13:09:18 & 28:56:30 & 58.67  & 85.66   & 4.00 &  $-28.91$ & - & 2.39 & - & 0.49  &
-   & 1.72 \\
33 & 13:13:42 & 28:35:30 &  50.78   & 84.86   & 3.86 &  $-26.28$ & $-28.05$ & 1.73 & 5.59 &
0.31  & 0.11   & 2.26 \\ 
34 & 13:17:54 & 29:40:30 &  58.11   & 83.64    & 5.41 &  $-24.45$ & $-31.00$ & 5.63 & 12.63 & 0.53  & 0.35   & 14.53 \\ 
35 & 13:26:38 & 06:13:30 & 326.32   & 67.48   & 8.46    &  14.27 & 19.92 & 4.60 & 10.52 & 0.07  & 0.17   &
4.21 \\ 
36 & 14:56:34 & 04:35:30 & 1.43  &  52.62   &  6.00 &  $-74.43$ & - & 13.83 & - & 0.25  & -   & 4.81 \\
37 & 15:07:06 & 07:32:30 & 7.89 & 52.38  &4.40 &  $-27.67$ & - & 5.10 & - & 0.27  &
-   & 1.99 \\
38 & 15:10:06 & 06:56:30 & 7.81 & 51.42   & 4.99 &  $-80.48$ & - & 14.77 & - & 0.27
& -   & 5.05 \\
39 & 15:11:50 & 06:11:30 &  7.24  & 50.63   &  5.92 &  $-18.66$ & $-19.93$ & 2.61 & 6.60 &
0.81  & 0.16   & 6.19 \\ 
40 & 15:15:54 & 10:56:30 & 14.51  & 52.41   & 4.74 & 34.98 & - & 6.73 & - & 0.44  & -   & 4.23 \\
41 & 15:26:30 & 27:39:30 &  42.97 & 55.72  & 6.71   & $-52.19$ & $-66.99$ & 8.66 & 15.82
& 0.08  & 0.44   & 14.94 \\ 
42 & 15:33:18 & 27:26:30 &  43.00  &  54.19  &  6.48 &  $-26.83$ & - & 3.13 & - & 0.35  & -   & 1.78 \\
43 & 16:19:46 & 07:29:30 & 21.24 & 36.99   & 4.11  & $-47.19$ & $-37.62$ & 6.68 & 10.88
& 0.11  & 0.35   & 8.94 \\ 
44 & 16:25:38 & 05:28:30 &  19.93  &  34.74  &  3.62  & $-30.15$ & $-36.95$ & 5.17 & 6.31 & 0.27  & 0.19   & 5.06 \\ 
45 & 16:39:50 & 08:02:30 &  24.66  & 32.85   & 6.20  & $-17.80$ & $-20.23$ & 3.37 & 8.97 & 0.68  & 0.26
  & 8.89 \\ 
46 & 16:47:38 & 05:13:30 & 22.80   & 29.82     & 5.87 &  53.01 & - & 14.86 & - & 0.53  & -   & 11.34 \\
47 & 16:51:10 & 25:10:30 & 45.20  & 36.69   &  7.48 &  $-34.79$ & $-42.44$ & 6.08 & 13.83
& 0.49  & 0.29   & 13.83 \\ 
48 & 21:27:06 & 25:40:30 & 75.48  &   $-17.83$ & 5.48 &  27.76 & - & 2.02 & - & 0.43  & -   & 1.66 \\
49 & 21:28:26 & 20:29:30 &  71.66  & $-21.58$   & 7.48 &  37.45 & - & 5.11 & - & 0.86  &
-   & 7.44 \\
50 & 21:32:30 & 25:11:30 &  76.00  &$-19.05$   & 4.74  & 24.53 & - & 1.64 & - & 1.55  &
-   & 4.24 \\
51 & 21:33:46 & 22:42:30 &  74.31  & $-20.98$   &  6.71 &  29.11 & - & 3.35 & - & 1.86  &
-   & 10.49 \\
52 & 21:34:14 & 25:27:30 &  76.49 &  $-19.14$  &  5.37 &  15.35 & - & 2.80 & - & 1.69  &
-   & 8.81 \\
53 & 21:40:58 & 19:55:30 &  73.39  &   $-24.11$  & 5.49 & 22.01 & 23.19 & 0.60 & 4.26 &
0.51  & 0.82   & 7.37 \\ 
54 & 21:41:22 & 31:23:30 & 82.02   & $-15.99$  &5.84 &  $-39.49$ & $-35.38$ & 2.48 & 4.94 & 0.32  & 0.73   & 8.63 \\ 
55 & 21:43:10 & 24:33:30 &  77.34  & $-21.21$  & 4.97 & 32.89 & - & 6.34 & - & 1.23  & -   & 11.39 \\
56 & 21:44:42 & 17:37:30 &  72.23 & $-26.34$   & 4.90 &  23.35 & - & 5.98 & - & 1.05  &
-   & 10.28 \\
\hline\end{tabular}
\end{center}
%\end{tiny}
\end{table*}

\begin{table*}
\addtocounter{table}{-1}
%\begin{tiny}
\begin{center}
\caption{The compact cloud sample with derived Gaussian parameters}
\label{t:table3}
\begin{tabular}{c c c c c c c c c c c c c}
\hline
\hline
    \multicolumn{1}{c}{\small Cloud} &
    \multicolumn{1}{c}{ RA} &
    \multicolumn{1}{c}{ Dec } &
    \multicolumn{1}{c}{ l } &
    \multicolumn{1}{c}{ b } &
    \multicolumn{1}{c}{ $\theta$ } &
    \multicolumn{1}{c}{v$^1_{LSR}$} &
    \multicolumn{1}{c}{v$^2_{LSR}$} &
    \multicolumn{1}{c}{$\Delta V^1$} &
    \multicolumn{1}{c}{$\Delta V^2$} &
    \multicolumn{1}{c}{ T$_b^{Pk_1}$ } &
    \multicolumn{1}{c}{ T$_b^{Pk_2}$ } &
    \multicolumn{1}{c}{ N${\rm{_{HI}}}$ } \\
{\small   No} & {\small (J2000)} & {\small (J2000)} & (deg) &  (deg) &($^\prime$) &{\small (kms$^{-1}$)}& {\small (kms$^{-1}$)}&{\small (kms$^{-1}$)} & {\small ( kms$^{-1}$ )}&(K) &(K)  & {\small{(10$^{18}$}}   \\
   & (h:m:s) & (d:m:s) & & & && & & & &  & {\small cm$^{-2}$)} \\
\hline
\hline
57 & 21:47:22 & 20:01:30 &  74.62  & $-25.12$   & 3.50 &  32.62 & - & 2.61 & - & 0.61  &-   & 2.65 \\
58 & 21:47:26 & 24:12:30 &   77.83 & $-22.14$  & 5.74 &  78.05 & - & 8.04 & - & 0.74  &
-   & 9.17 \\
57 & 21:58:02 & 35:29:30 & 87.53 & $-15.19$  &  7.00 &  19.27 & - & 3.42 & - & 1.31  &
-   & 7.18 \\
60 & 21:59:02 & 22:16:30 &  78.52 & $-25.37$  & 6.06 &  71.41 & - & 6.57 & - & 0.75  &
-   & 6.51 \\
61 & 21:59:30 & 34:50:30 &   87.35  & $-15.89$     &  6.11  & 29.98 & 25.03 & 3.74 & 3.99 &
1.05  & 0.71   & 13.22 \\ 
62 & 21:59:50 & 24:01:30 &  79.96  &  $-24.19$     & 4.24   & 23.11
& - & 2.72 & - & 1.38  & -   & 6.91 \\
63 & 22:00:58 & 23:46:30 &   79.99  &  $-24.55$   &  5.46 &  22.91 & - & 3.26 & - & 1.23  &
-   & 6.79 \\
64 & 22:01:06 & 34:50:30 &  87.61  &  $-16.09$    &  6.48 &  24.79 & - & 3.71 & - & 1.17  &
-   & 7.71 \\
65 & 22:02:34 & 16:53:30 &  75.06  & $-29.88$   & 6.48 &  21.98 & - & 4.50 & - & 0.95  &
-   & 7.52 \\
66 & 22:03:26 & 17:37:30 &  75.82 &   $-29.48$   &   5.24 &  49.37 & - & 7.20 & - & 1.09  &
-   & 11.43 \\
67 & 22:04:18 & 20:37:30 &  78.30 &  $-27.41$ & 5.20 &  33.11 & - & 6.36 & - & 0.92  &
-   & 8.92 \\
68 & 22:04:34 & 21:12:30 &  78.79 &  $-27.02$  &  6.34 &  39.01 & 37.39 & 3.77 & 8.82 &
1.58  & 0.39   & 18.43 \\ 
69 & 22:04:54 & 14:57:30 &  73.98 & $-31.66$ & 4.12 &  58.52 & - & 5.62 & - & 0.31  & -   & 2.67 \\
70 & 22:06:18 & 04:58:30 &  65.41 & $-38.80$ & 7.65 &  $-47.42$ & $-46.84$ & 2.62 & 12.50
& 0.48  & 0.38   & 11.77 \\ 
71 & 22:13:34 & 16:22:30 &   76.91 & $-32.04$  & 5.11  & 54.28 & - & 4.23 & - & 0.91  & -   & 6.67 \\
72 & 22:17:34 & 21:49:30 &  81.84 & $-28.51$  & 6.24 &  $-30.72$ & - & 2.36 & - & 1.13  &
-   & 4.86 \\
73 & 22:18:22 & 21:41:30 &  81.91 & $-28.73$  & 5.24   & $-34.72$ & - & 2.39 & - & 1.19  &
-   & 4.82 \\
74 & 22:12:54 & 25:20:30 &  83.38 & $-25.10$ &  5.74 &  $-69.86$ & $-71.43$ & 4.17 & 18.09
& 0.41  & 0.37   & 16.46 \\ 
75 & 22:39:50 & 13:11:30 &  80.36 & $-38.56$  & 5.24 &  $-32.76$ & - & 2.19 & - & 0.83  & -   & 3.34 \\
76 & 22:50:06 & 34:42:30 & 96.36   & $-21.82$   &  8.66 &  12.66 & 13.72 & 1.84 & 3.59 &
2.50  & 0.51   & 12.56 \\ 
77 & 23:11:26 & 27:58:30 &  97.42  & $-29.89$  & 3.50 &  75.52 & - & 4.49 & - & 0.59  & -   & 3.10 \\
78 & 23:14:14 & 25:47:30 &  96.98 &  $-32.13$  &  4.74 & 42.57 & - & 2.01 & - & 0.97  &
-   & 3.74 \\
79 & 23:22:42 & 13:35:30 & 92.28 &  $-43.93$ & 3.74   & $-63.10$ & $-62.16$ & 1.93 & 3.89 &
0.22  & 0.15   & 1.99 \\ 
80 & 23:26:54 & 30:00:30 & 101.93  & $-29.39$  & 4.97 & $-37.63$ & $-35.95$ & 1.99 & 2.87 & 0.63  & 0.21   & 3.99 \\ 
81 & 22:54:46 & 08:02:30 &  80.00 &  $-44.88$  &  4.08 & $-60.13$ & - & 9.22 & - & 0.44  & -   & 5.41 \\
82 & 22:57:30 & 32:24:30 &  96.65 & $-24.58$  & 5.74 &  16.19 & 18.39 & 2.20 & 4.91 &
1.41  & 0.23   & 8.21 \\ 
83 & 23:00:54 & 32:35:30 &  97.44 & $-24.75$  & 7.12 &  15.39 & - & 1.88 & - & 4.13  & -   & 15.14 \\
84 & 23:01:58 & 32:24:30 &  97.57  & $-25.01$   & 5.74 &  18.12 & 18.24 & 1.87 & 3.28 &
4.45  & 1.21   & 23.96 \\ 
85 & 23:06:50 & 30:29:30 & 97.65  &  $-27.19$ & 4.50  & 39.95 & 40.17 & 2.39 & 12.86 & 0.71  & 0.46   & 14.77 \\ 
86 & 23:10:30 & 31:01:30 &  98.71  & $-27.05$  & 8.0 &  $-24.01$ & - & 1.91 & - & 1.26  & -   & 4.56 \\
87 & 23:18:06 & 30:39:30 &  100.21 & $-28.06$  & 6.48 &  24.58 & 24.28 & 2.61 & 7.16 &
1.02  & 0.51   & 12.26 \\ 
88 & 23:23:18 & 35:51:30 & 103.54  & $-23.67$   &  6.87 &  $-53.14$ & - & 4.33 & - & 2.61  & -   & 20.74 \\
89 & 23:29:14 & 22:00:30 &  98.88 & $-36.99$  & 5.59 &  8.97 & - & 1.45 & - & 1.64  & - & 3.81 \\
90 & 23:34:38 & 22:10:30 &  100.41 & $-37.31$  & 6.15 &  $-21.74$ & - & 3.74 & - & 1.44  & -   & 8.49 \\
91 & 23:36:02 & 13:41:30 &  96.44 & $-45.26$  &  4.74   & $-64.30$ & - & 1.93 & - & 1.23  &-   & 4.66 \\
92 & 23:36:14 & 22:34:30 &  101.43 & $-37.07$   &  5.98 &  13.90 & - & 1.91 & - & 2.40  &
-   & 8.52 \\
93 & 23:39:02 & 13:36:30 & 97.34  & $-45.63$  &  6.71  & $-54.89$ & $-56.07$ & 2.18 & 5.43 &
2.22  & 0.24   & 11.98 \\ 
94 & 23:42:50 & 21:36:30 &  102.42 & $-38.51$  &  7.35 &  10.61 & - & 2.53 & - & 3.02  & -   & 13.59 \\
95 & 23:43:26 & 11:53:30 & 97.80 & $-47.63$  &  4.00   & $-47.34$ & - & 2.54 & - & 0.83  &
-   & 3.51 \\
96 & 23:43:30 & 11:46:30 & 97.76  & $-47.74$  &  5.00   & $-52.54$ & - & 3.03 & - & 2.45  &
-   & 13.64 \\
\hline\end{tabular}
\end{center}
%\end{tiny}
\end{table*}

\begin{figure*}
\epsscale{2.0}
\plotone{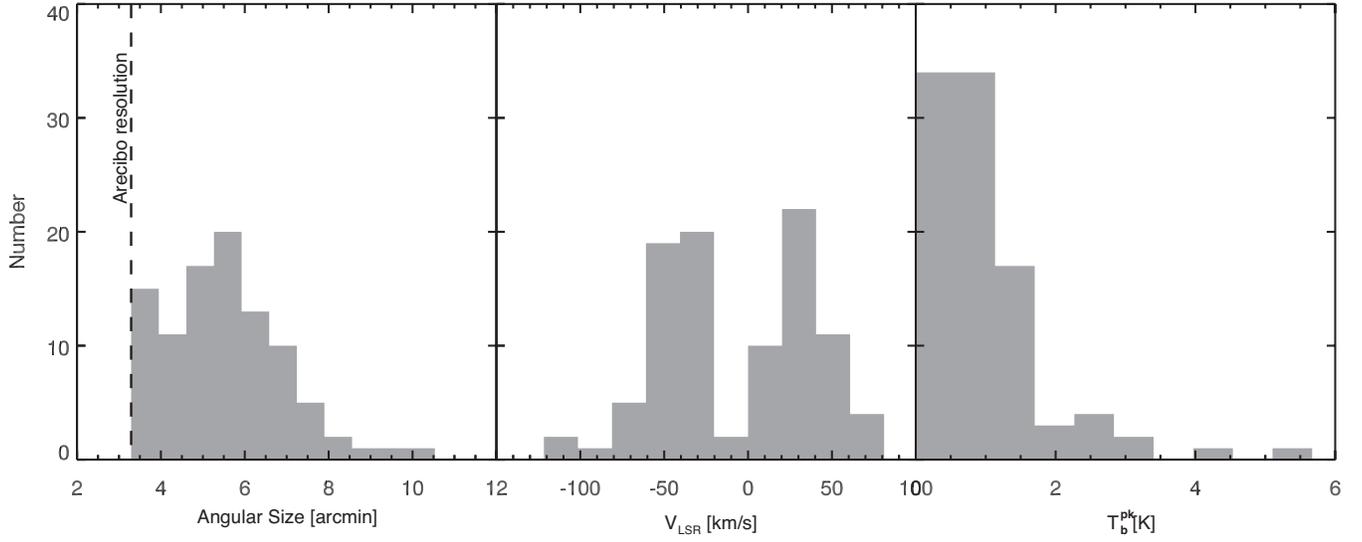}
\caption{\label{f:hist_1} Histograms of basic observed properties for
the whole sample of compact clouds: angular size, peak V$_{LSR}$,
and the peak
brightness temperature, T$_b^{pk}$. }
\end{figure*}

\begin{figure*}
\epsscale{0.5}
\includegraphics[scale=0.9,angle=0.0]{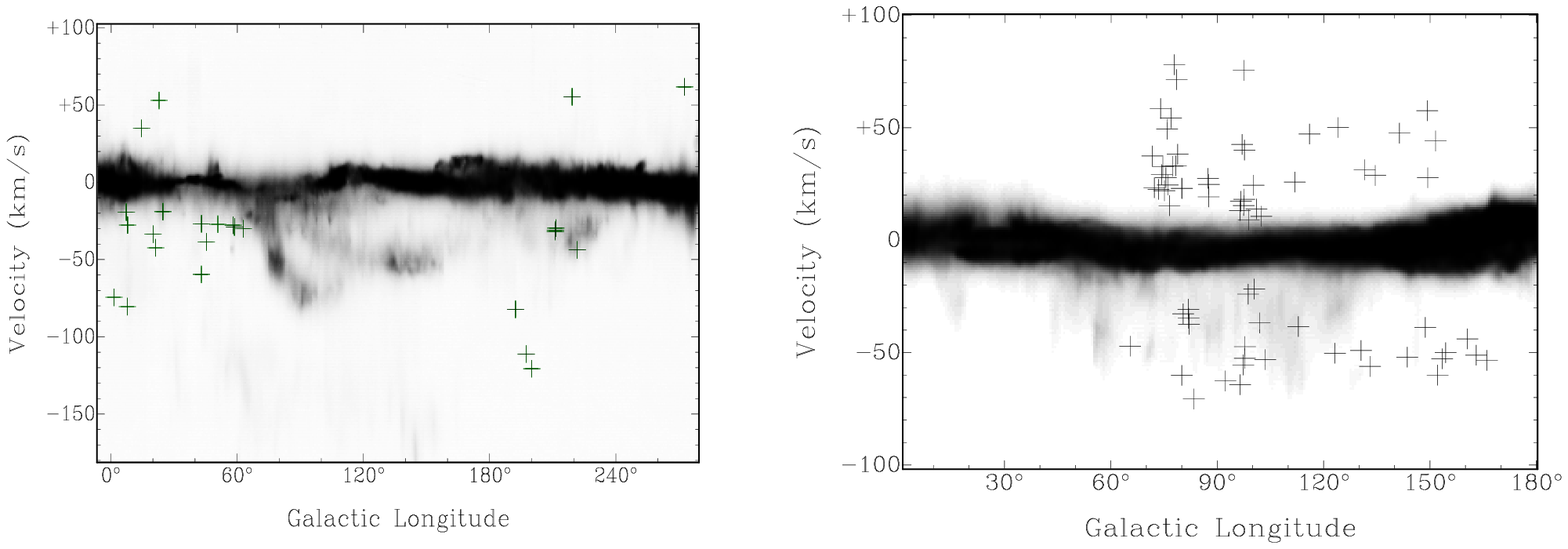}
\caption{\label{f:l_v} {(\it{Left})} Position-velocity diagram for the Milky Way
centered at b=40$^\circ$ and integrated over the range
b=[30$^\circ$:50$^\circ$], using data from the LAB survey. The
positions of compact clouds with $|b|>$0$^\circ$ are plotted as crosses.
{(\it{Right})} Position-velocity diagram for the Milky Way
centered at b=$-$40$^\circ$ and integrated over the range
b=[$-$50$^\circ$:$-$30$^\circ$], using data from the LAB survey. The
positions of compact clouds with $|b|<$0$^\circ$ are plotted as crosses.
}
\end{figure*}

\subsection{Completeness Test}
\label{ssec:complete}

In order to check the completeness of our cloud catalog,
we injected fake compact clouds in several of the large survey data
cubes (40 degree $\times$ 18 degree $\times$  240 \kms~size). The fake
clouds were injected at random locations with input parameters,
viz.~LSR velocities; FWHM; amplitudes; angular sizes; aspect ratios;
and position angles, chosen randomly from a range
determined by the observed cloud properties.
Further, as suggested by observations, both the
velocity profile and the spatial distribution of each fake cloud were
modeled with Gaussian functions. Four different simulations were
tried, each sampling a different range of
FWHM and peak brightness temperature of the compact clouds:

\begin{itemize} \item
{\bf{Simulation A}}:  Bright clouds with a narrow
velocity linewidth: T$^{pk}_b = 1.2-3.2$ K, FWHM $= 2.0-6.0$ km~s$^{-1}$.

\item
{\bf{Simulation B}}: Bright clouds with a wide velocity linewidth:
T$^{pk}_b = 1.2-3.2$ K, FWHM $= 6.0-12.0$ km~s$^{-1}$.

\item
{\bf{Simulation C}}: Faint clouds with  a narrow velocity linewidth:
T$^{pk}_b = 0.2-1.2$ K,  FWHM $= 2-6$ km~s$^{-1}$.

\item
{\bf{Simulation D}}: Faint clouds with a wide velocity linewidth:
T$^{pk}_b = 0.2-1.2$ K, FWHM $=6-12$ \kms. \end{itemize}

The range of ${\rm{V_{LSR}}}$ over which clouds were injected  was kept the same
for all four simulations, as $-120 \le {\rm{V_{LSR}}} \le +120$ \kms.
Similarly, the cloud angular size, position angle and aspect ratio
were in the range 3.0--8.0 arcmins, 0--180 degrees, and 1.0--2.0,
respectively. For each simulation, 15 fake clouds were injected. The
input parameters of the injected clouds were noted
to be later compared with the list of recovered clouds.

The data cube with injected fake clouds was then searched by eye
following the same criteria adopted for the selection of ``real"
compact cloud as discussed in Section~\ref{sec:search}.
The list of detected compact clouds was compared with the list of
injected fake clouds to compute the fraction of fake clouds  missed by
our search.  Table~\ref{tab:fake} summarizes the results
of our completeness test. Column (1) shows the simulation type, and
column (2) the detection rate for the fake clouds. Column (3) shows
the fraction of fake clouds with V${\rm{_{LSR}}}\sim$ 0 \kms
; compact clouds with V${\rm{_{LSR}}}\sim$ 0 \kms~were missed in the search
because of confusion with bright Galactic emission. Column (4) shows
the fraction of fake clouds which appeared to be spatially connected
to some large-scale emission.
%hence were missed as they did not satisfy our selection criterion.
Column (5) shows the fraction of detected real clouds in our catalog,
with properties similar to the input parameters
of the simulations.

As fake clouds were injected at random locations in the data cubes,
with V${\rm{_{LSR}}}$ selected randomly in the specified range of $-120$ to
$120$ \kms, some of the fake clouds have  V${\rm{_{LSR}}} \sim 0$ \kms. Such
clouds were missed by our search as velocity channels around V${\rm{_{LSR}}}$
$\sim 0$ \kms~are dominated by bright Galactic emission.
Quantitatively, we find that we cannot recover any fake clouds placed
in regions where the Galactic background (as determined from the
Leiden/Argentine/Bonn survey, \cite{lab}) exceeds 4 K, but
that in regions below 4 K our recovery rate does not substantially
depend on the Galactic background emission. 
This is consistent with our finding that few real clouds were detected
with V${\rm{_{LSR}}}$ close to 0 \kms~(see Figure~\ref{f:hist_1}). In
addition, some of the fake clouds were missed as they coincided
spatially with large-scale structures, hence they were rejected by our
selection criteria.
Taking into consideration the above two cases, we find a detection
rate of 100\% for the clouds which are bright
and have either narrow or wide velocity linewidths. However when the
clouds are faint,  their detection rate decreases to
80\% in the case of narrow velocity linewidths, and 70\% in the case of
broad velocity linewidths.

Comparing the fake cloud detection rate with the real cloud detections
suggests that
(i) a lack of  bright compact clouds with wide velocity linewidths in
our catalog is a real feature and not a selection effect; (ii) it is
unlikely that we have missed any bright
clouds with narrow velocity widths in the regions where Galactic
background does not exceed 4 K; and (iii) the observed decrease in the
number of detections of faint compact clouds with a wide linewidth is
likely to be a selection bias.
To summarize, in our selected region of study ($\sim$4600 deg$^2$), we
cannot recover any compact clouds in regions where the Galactic
background exceeds 4 K. On the other hand, in regions with Galactic
background below 4 K, we are most likely missing about one quarter of
the clouds, in total, due to our selection biases.

\begin{table*}
\caption{Results of fake clouds simulations}
\centering
\label{tab:fake}
\begin{tabular}{c c c c c}
\hline
\hline
    \multicolumn{1}{c}{Simulation} &
    \multicolumn{1}{c}{Detection rate} &
    \multicolumn{1}{c}{Confusion} &
    \multicolumn{1}{c}{Connected} &
    \multicolumn{1}{c}{Observations} \\
\hline
Case A &13/15 & 0  & 2/15 & 22/96\\
Case B & 10/15 & 3/15  & 2/15 & 0/96\\
Case C & 8/15 &0 & 2/15 & 52/96\\
Case D & 9/15& 2/15 & 1/15 & 22/96\\
\hline\end{tabular}
\end{table*}

\section{Observed cloud properties}
\label{sec:properties}

\subsection{Clouds size, peak brightness temperature, and central velocity}

Figure~\ref{f:hist_1} shows histograms of basic observed properties
for the whole compact cloud sample: angular size, peak LSR velocity,
V$_{\rm{LSR}}$, and peak brightness temperature, ${\rm {T^{pk}_b}}$.

The clouds are typically very compact, with the median angular size of
the sample being $\sim 5'$. Many clouds are unresolved at Arecibo's
resolution, at least in one dimension. A large fraction of the sample
has an unresolved core along   with some faint diffuse emission, as
seen in
Figure~\ref{f:image1}. The extent of this faint diffuse emission is
most likely slightly overestimated (by about an arcmin) due to the
telescope sidelobes.

The majority of clouds have ${\rm { T^{pk}_b}}=0.5-2$ K, while the median peak
brightness temperature
for the whole sample is 0.75 K.
This low peak brightness temperature, coupled with the small angular
size, explains why such clouds were largely missed by previous
large-scale Galactic HI surveys, e.g. the GASS and LAB surveys
\citep{gass,lab}.

Clouds are found at both positive and negative LSR velocities. The
$V_{\rm LSR}$ histogram has a nearly symmetric distribution around
$V_{\rm LSR}=0$ \kms.
A lack of clouds  between  $-20 \leq V_{\rm LSR} \leq 5$ km~s$^{-1}$ is
not real, but due to difficulties in finding clouds in the presence of
bright Galactic emission, as discussed in  Section
\ref{ssec:complete}. It is interesting to note that if we neglect bins
$|V_{\rm LSR}| <20$ \kms, a Gaussian function could be fit
with a peak at $\sim$ 0 \kms~and a standard deviation of $\sim$ 42 \kms.

As observed clouds have typically $|b| \sim 30^\circ-50^\circ$ (see Table~\ref{t:table3}),
Figure~\ref{f:l_v} shows the $l$ vs V$_{\rm LSR}$  plots of the LAB survey
integrated over $30^\circ < b < 50^\circ$ ({\it{left}}) and $-50^\circ < b < -30^\circ$ ({\it{right}}),
with the positions of compact clouds with $b> 0^\circ$ and $b<0^\circ$ overplotted, respectively. 
Most clouds are found in the 1st and 2nd Galactic quadrant, at least
partially due to our limited survey coverage, and appear at velocities
both allowed and forbidden by Galactic rotation. %The number of clouds
%at allowed vs forbidden velocities is similar. 
%In addition, no obvious trend with the terminal velocity is noticeable.
%This suggests that cloud motions are not dominated by Galactic rotation. This could happen if 
%clouds are very distant, or very local.
To quantify the extent to which cloud velocities deviate from the velocity
expected from Galactic rotation, we define the
deviation velocity (Wakker 1991) as:
\begin{equation}
{\rm{V_{dev}}}= \left\{ \begin{array}{ll}
{\rm{V_{LSR}-V_{min}}}(l,b) & if ~{\rm {V_{LSR}}}<0 \\
{\rm{V_{LSR}-V_{max}}}(l,b) & if ~{\rm{V_{LSR}}}>0
\end{array}
\right .
\end{equation}
 where (V${\rm{_{min}}}$, V$_{\rm max}$) is the range of velocities allowed by
Galactic rotation for a given $(l,b)$ direction. For clouds
following Galactic rotation, V$_{\rm dev}=0$. Considering that the above
definition of V$_{\rm dev}$ does not account
for the velocity dispersion due to random and turbulent motions,
clouds with ${\rm{|V_{dev}}}|<20-30$ \kms~are often considered as not being
at forbidden velocities.
Traditionally, high velocity clouds (HVCs) are assumed to have
${\rm{|V_{dev}}}|> 60$ \kms~and
${\rm{|V_{LSR}}}|>90.0$ \kms, while intermediate velocity clouds (IVCs)
have ${\rm{|V_{LSR}}}|= 40-90$ \kms~\citep{wakker04}.

\begin{figure*}
\epsscale{1.}
\plotone{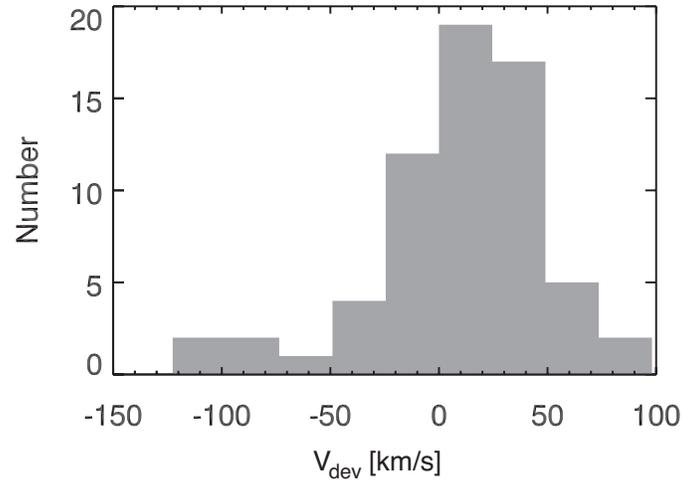}
\caption{\label{f:Vdev} Histogram of the deviation velocity for the compact
clouds population. Only clouds with non-zero ${\rm{V_{dev}}}$ have been
included in the plot.
}
\end{figure*}

We find that $\sim$33\% (32/96) of the sample clouds have V$_{\rm dev}
\sim0.0$, i.e.~velocities allowed by Galactic rotation.
For the remaining clouds with non-zero V$_{\rm dev}$,  most of the clouds
deviate from
Galactic rotation  by ${\rm{|V_{dev}}}| \leq 50$  \kms~(with a majority
showing ${\rm{V_{dev}}} \sim$ 20 km~s$^{-1}$). Figure~\ref{f:Vdev} shows the
histogram of deviation velocities for the whole catalog; clouds with
${\rm{V_{dev}}}\sim 0$ \kms~have not been included in this graph.
 We find that only two clouds in our
sample (Cloud No 27, 28) have ${\rm{|V_{dev}}}|> 60$ \kms~and V$_{\rm LSR}>90.0$
\kms, hence can be classified as HVCs. Similarly, only a few compact
clouds
can be classified as IVCs.

\subsection{Clouds FWHM, N${\rm{_{HI}}}$, and T$_{K, max}$}

\begin{figure*}
\epsscale{2.}
\plotone{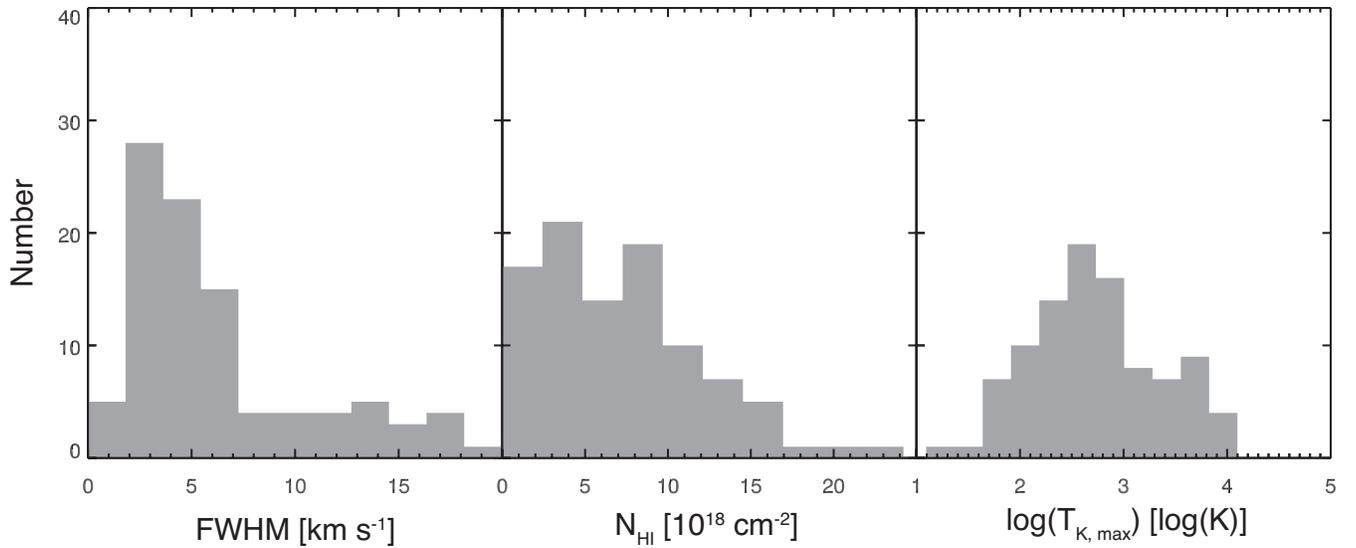}
\caption{\label{f:temp} Histograms showing distributions of FWHM,
N${\rm{_{HI}}}$ and log(Kinetic temperature)
for the compact cloud sample.
}
\end{figure*}

Figure~\ref{f:temp} shows histograms of the velocity FWHM ($\Delta
V$), the HI column density (${\rm{N_{HI}}}$), and the upper limit on the
kinetic temperature (i.e.~the kinetic temperature
in the case of no non-thermal broadening, defined as T$_{K, max}=21.86
\Delta V^2$).
As discussed in the next sub-section, some clouds have multi-Gaussian
velocity profiles, and
for such clouds we include only the brightest component when
considering the above quantities.

Generally, the FWHM  for the sample is in the range of $\sim$ 1$-$8
\kms,  with a median FWHM of 4.2 \kms. 
Most of the clouds are cold (in cases
where the velocity profiles were fitted with single Gaussian) or have
a cold core (in the case of double Gaussian fits).
  The median T$_{K, max}$  of the sample is only 300 K.
In their HI absorption survey \cite{Heiles03b} found the CNM spin 
temperature of 50-70 K (for $|b|>10$ degrees) and a typical sonic
Mach number of about 3. This results in a typical CNM kinetic temperature
$T_{k, max}\sim150-250$ K. Our median
$T_{\rm k, max}=300$ K is very similar to what is
found for the CNM clouds seen in absorption, suggesting that
these compact clouds have properties similar to those of typical
Galactic CNM clouds.

The integrated HI column density distribution peaks at $5 \times
10^{18}$ cm$^{-2}$, with $\sim$ 80\%  of clouds having
N${\rm{_{HI}}}<10^{19}$ cm$^{-2}$. This is low relative to typical Galactic
CNM clouds which typically have N${\rm{_{HI}}}>10^{19}$ cm$^{-2}$.
Our HI column density was derived assuming the gas to be optically
thin, an assumption that is
reasonable because the emission is faint. As many compact
clouds are unresolved or only
 marginally resolved with the Arecibo, the beam dilution effect is likely
to be important for the smallest clouds, and
 hence the derived N$\rm{_{HI}}$ could be lower than the true
${\rm{N_{HI}}}$. For example, high resolution interferometric follow-up 
observations (Peek et al, in preparation) 
of some of the compact clouds unresolved with 
Arecibo revealed bright compact cores
(sizes below the resolution of 28 arcsec) embedded in a diffuse emission, 
with a peak ${\rm{N_{HI}}}$ $\sim$10 times the value seen by Arecibo.

\begin{figure*}
\epsscale{1.}
\plotone{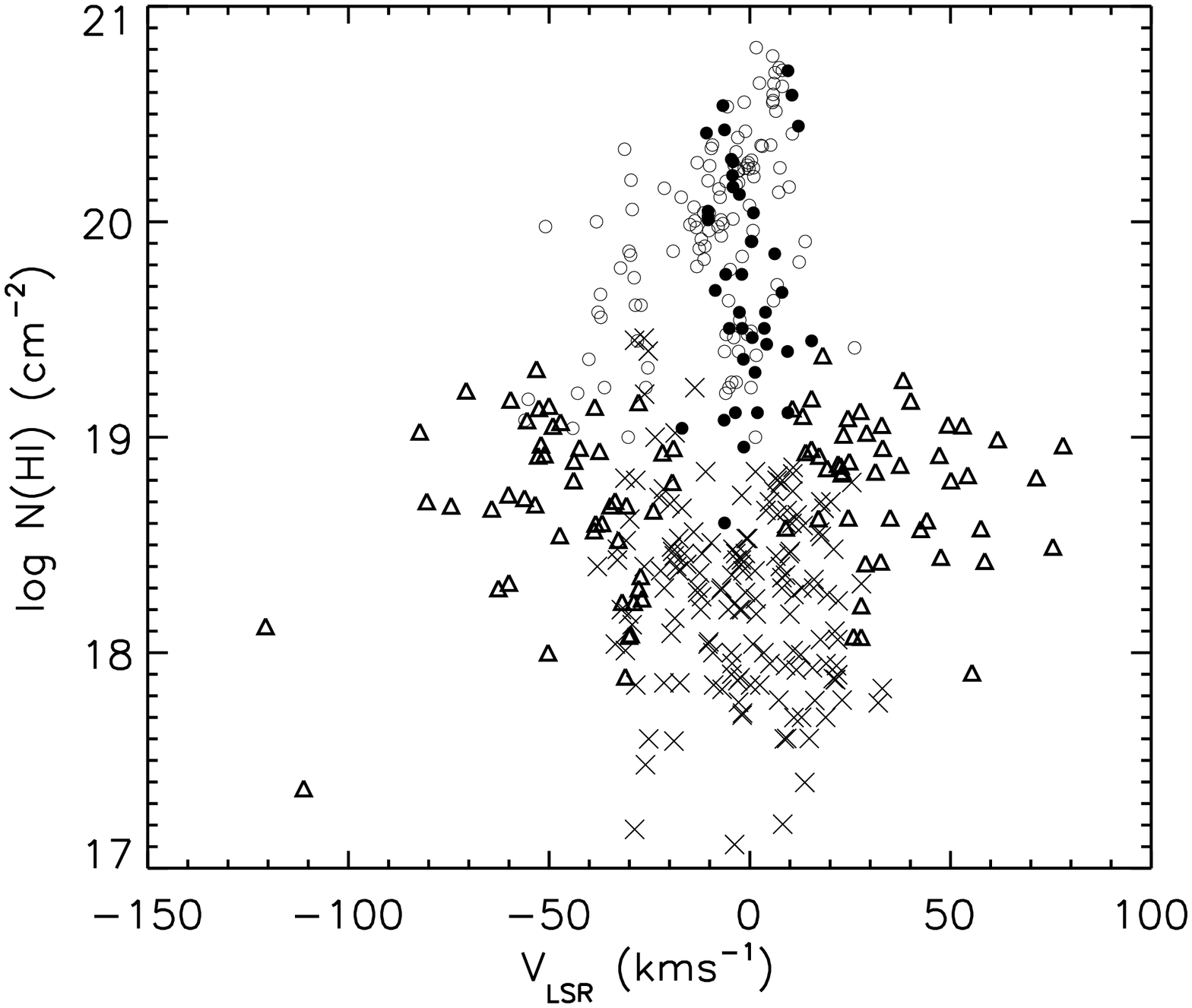}
\caption{\label{f:lism} Velocity distribution of compact clouds (shown
as triangles) along with LISM clouds (shown as crosses). The LISM
clouds are towards stars within 50 pc (\cite{muller06});
the data for LISM clouds is taken from \cite{frisch02},\cite{red02}
and \cite{wood05}. Also shown
in the plots are the CNM (solid circle)  and WNM (empty circle)
Galactic clouds from
\cite{Heiles03a}. Only sight lines  with  $\left| b \right|
\geq25^\circ$ are considered. }
\end{figure*}

%The integrated HI column density distribution peaks at $5 \times
%10^{18}$ cm$^{-2}$, with $\sim$ 80\%  of clouds having
%N${\rm{_{HI}}}<10^{19}$ cm$^{-2}$. This is low relative to typical Galactic
%CNM clouds.
%This is similar to the HI column density of the population %of mini
%high velocity clouds (mini-HVCs)
%which was found to peak at $3.0 \times 10^{18}$ cm$^{-2}$ (\cite{hoffman02}).
In Figure~\ref{f:lism} we compare our compact clouds with typical
Galactic CNM and WNM clouds, and also with
interstellar clouds found in the local ISM (LISM).
The corresponding HI column density and ${\rm{V_{LSR}}}$ of LISM clouds were
compiled by \cite{muller06}
and originate from HI, D$^0$ and Ca$^+$ observations toward stars
within 50 pc. The same quantities are also plotted for the CNM and WNM
Galactic clouds from the \cite{Heiles03a} survey,
for high latitude line of sights ($\left| b \right| \geq25^\circ$).
LISM clouds (shown with crosses) have a distribution symmetric about V${\rm{_{LSR}}}\sim0$
\kms,  and ${\rm{N_{HI}}}\sim 2 \times 10^{18}$ cm$^{-2}$. This is
significantly lower than
what is measured for the CNM/WNM, ${\rm{N_{HI}}}> {\rm a~few~} \times
10^{19}$ cm$^{-2}$ (shown with circles).
The HI column density of our compact clouds (shown with triangles),
even after taking into account the beam dilution effect, fills in the gap
between LISM clouds and the CNM/WNM clouds. The velocity distribution
of compact clouds is however more extended than that of the other two
populations. The gap in the distribution of compact clouds around
${\rm{V_{LSR}}}\sim0$ \kms~is not real, but due to the presence of strong
Galactic emission.

Another population of Galactic CNM clouds with similarly low HI column
density was discovered recently by
\cite{bk05} and \cite{snez05} through
HI absorption measurements.
These so called `low column density' absorbers have
a peak optical depth of $10^{-2}$, a FWHM of 3 \kms, and a typical HI
column density
${\rm{N_{HI}}}\sim 6 \times 10^{18}$ cm$^{-2}$ (\cite{snez07}).
In fact, Braun \& Kanekar (2005) observed several compact HI blobs in HI
emission near one such absorber, with angular sizes of only $\sim 2-3$
arcmin.

\subsection{HI line profile shapes: evidence for a multiphase medium}

\begin{figure*}
\epsscale{2.0}
\plotone{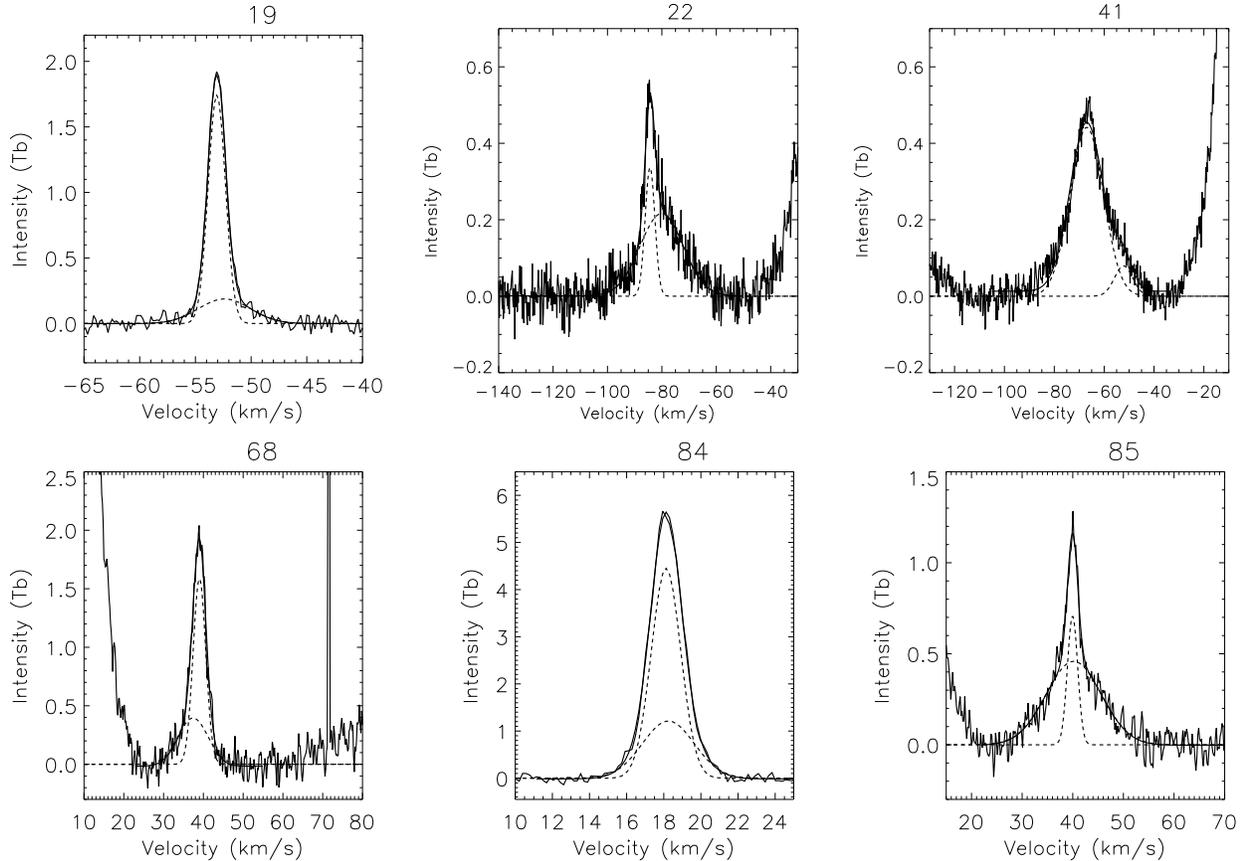}
\caption{\label{f:profile} Several examples of HI line profiles fitted
with two Gaussian
functions. Two Gaussian components are shown with dashed lines. }
\end{figure*}

For each compact cloud, the integrated HI line profile was first
fitted with a single Gaussian component and the residuals were
inspected. HI profiles in some cases were found to deviate
significantly from a simple Gaussian function. Such profiles were then
fitted with a greater number of Gaussian components.
We found that in 66/96 ($\sim69$\%) cases a single Gaussian function
provided a good representation to the line profiles,  whereas in 30/96
(31\%)  two Gaussian functions were required to fit the observed
profiles.  Examples of some shapes of line
profiles  fitted with a double Gaussian component are provided in
Figure~\ref{f:profile}.

For some clouds with double Gaussian components (e.g.~Clouds No 84 and
85 in Figure~\ref{f:profile}), velocity profiles show evidence for a
core-halo structure with two spectral components. A narrow and bright
Gaussian function is required to fit the line center, while a faint
and broad component is  needed to fit the line wings. Such cases could
indicate  the existence of a multi-phase medium. However, in most
cases with double Gaussian fits (17 out of 30 clouds), line profiles
show weak, broad velocity tails, e.g.~V$_{\rm LSR}\sim [-80,-60]$ \kms~for
cloud No 22 in Figure~\ref{f:profile}. In such cases, the narrow
Gaussian component is significantly off-set in velocity from the broad
Gaussian  component. Such  profiles with weak velocity tails are
likely to be due to the contribution of diffuse Galactic background
emission, and/or
superposition of different clouds along the line of sight.
Hence, for computing the physical parameters for clouds with multiple
Gaussian fits, only the brighter Gaussian component was considered for
the rest of the analysis.
% or a superposition of multiple clouds along %the line of sight

\begin{figure*}
\epsscale{0.8}
\plotone{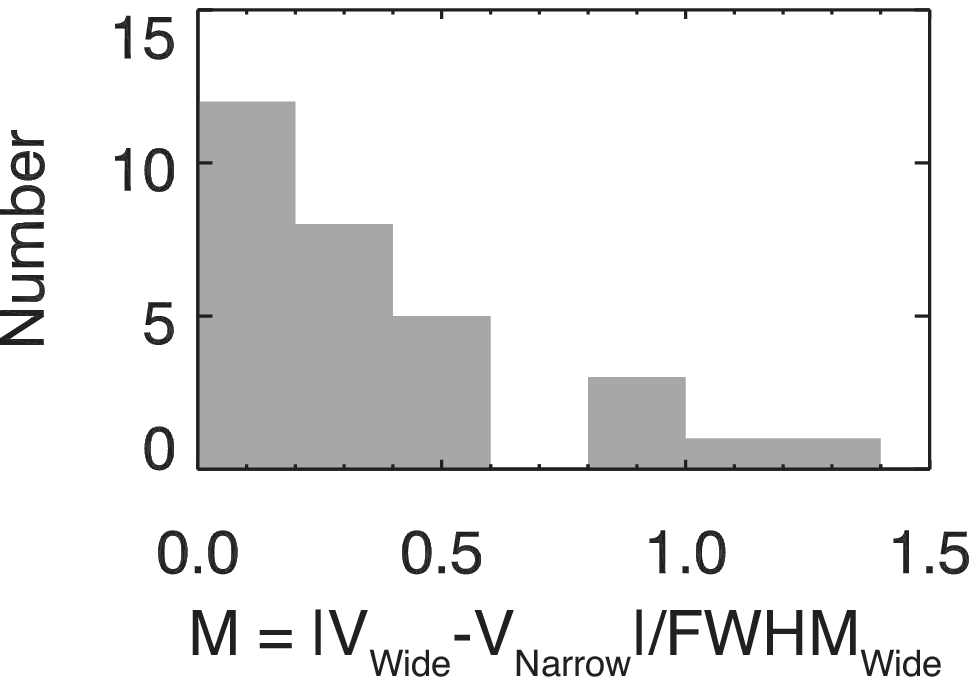}
\caption{\label{f:macno} Histogram showing the ratio of the absolute
difference between the centers of the two fitted Gaussian functions to
the FWHM of the broader function, for the clouds whose velocity
profiles are represented by two Gaussian functions. }
\end{figure*}

In order to characterize  motions of the colder cores (represented by
narrow Gaussian functions) relative to the warmer envelopes
(represented by broad Gaussians) for those clouds with velocity
profiles well fit by two Gaussian functions, we estimated the sonic
Mach number ({\bf{M}}) of the cores.
Following \cite{kh06}, we define {\bf{M}}  as the the ratio of the
absolute difference between the centers of
the two fitted Gaussian functions to the FWHM of the broader function:
\begin{equation}
{\bf{M}}={\rm{\frac{|V_{Wide}-V_{Narrow}|}{FWHM_{Wide}}}}.
\end{equation}
The {\bf{M}} histogram in Figure~\ref{f:macno} shows a peak at ${\bf{M}}=0.1$
and a tail all the way to
${\bf{M}}=1.5$. About half of the multi-phase clouds have ${\bf{M}}\la0.2$ and 2/3
of the multi-phase sample has ${\bf{M}}\la0.4$. Clouds with ${\bf{M}}\ga0.2$ show
velocity tails.
Clouds with ${\bf{M}}\la0.2$, which are likely to have real multi-phase signatures,
have cold cores with low Mach numbers, indicating that the cold cores
are moving subsonically within warmer envelopes. For comparison,
\cite{Heiles03b} found supersonic internal motions for Galactic CNM
clouds with ${\bf{M}}\sim3$,
while Kalberla \& Haud (2006) found that most HVCs have ${\bf{M}}\sim1.5$ for
cold cores relative to their warm envelopes. This may suggest that
compact clouds are less turbulent than Galactic CNM clouds or HVCs.
However, we are most likely underestimating {\bf{M}} due to
our inability to spatially resolve cold cores.

\subsection{Compact clouds with velocity gradients}
\label{ssec:velgrad}

\begin{figure*}
\epsscale{2.2}
\plotone{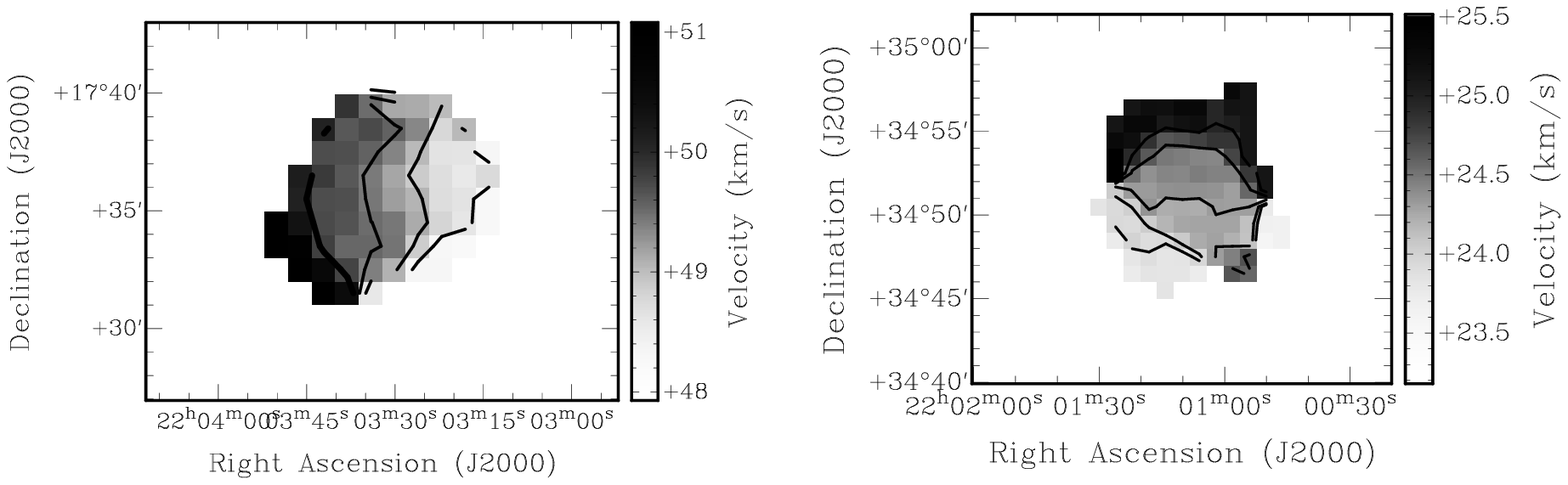}
\caption{\label{f:mom1} The HI velocity field for two compact clouds
with significant velocity gradient. {{\bf(Left)}} Cloud 66: The
contours are in steps of 0.5 kms$^{-1}$,
starting from 48.5 \kms. {\bf{(Right)}} Cloud 64: The contours are in steps of
0.3 kms$^{-1}$, starting from 23.7 kms$^{-1}$.
}
\end{figure*}

The high spatial and velocity resolution of the GALFA-HI survey allows us to
investigate velocity gradients across HI clouds.
We find that 31 clouds in our sample (i.e 32\% of the clouds) show
velocity gradients. To map the velocity gradient we derived the
velocity centroid at each position in the cloud, using the
intensity-weighted
first moment map. Before obtaining the moment map, lines of sight with
a low signal to noise ratio were excluded by applying a cutoff at the
$2\sigma$ level,
$\sigma$ being the rms noise level in a line free channel. This was
done after the smoothing of data cubes in velocity (using boxcar
smoothing three channels wide) and position (using a Gaussian with
${\rm FWHM} \sim 1.5$ times that of the Arecibo beam). Further, for
compact clouds
with velocities close to Galactic
emission,  each velocity channel was inspected and pixels with
Galactic emission were blanked before obtaining the velocity field.

\begin{figure*}
\epsscale{1.}
\plotone{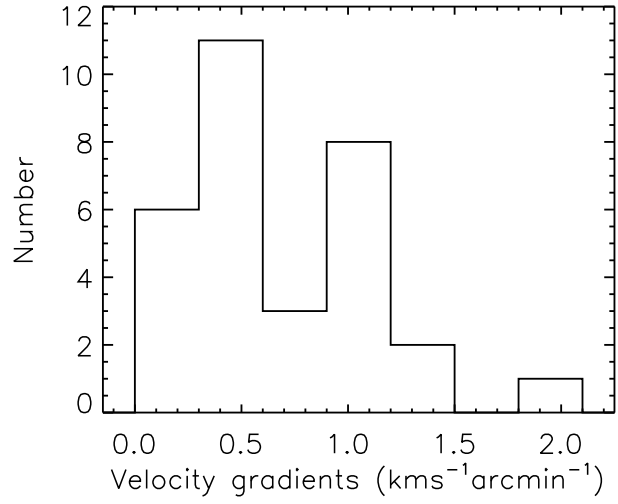}
\caption{\label{f:v_grad} Histogram of the velocity gradient per unit
angular size for the 31 compact clouds with a velocity gradient. }
\end{figure*}

Figure~\ref{f:mom1} shows two example velocity fields, while
Figure~\ref{f:v_grad}  shows the histogram of measured velocity
gradients, along the kinematic major axis. We find typically a
velocity gradient of $\sim 0.5 - 1$ km~s$^{-1}$~arcmin$^{-1}$.
However, since the clouds are mainly unresolved,  the observed
velocity gradient could be significantly affected by beam smearing;
beam smearing leads to an increase in the size of the cloud and a
decrease in the observed velocity gradient. Our measured velocity
gradient is therefore most likely underestimated.

One possibility is that the velocity gradients arise due to
fluctuations in the background HI emission, especially for the clouds
in regions with complicated Galactic emission. We note however, that
all of the clouds  with systematic gradients are found to be very
isolated with no strong background Galactic emission.
Further, for each cloud with a velocity gradient, the line profiles
were carefully inspected to check for baseline stability. No
systematic change in the baseline shape was found.
These tests suggest that the measured velocity gradients are inherent
to the clouds.

A large-scale velocity gradient could be a signature of rotation in
the case of self-gravitating clouds. For example, velocity gradients have
been noted in many CO studies of molecular clouds in the Milky Way and
in M33 (Philips 1999, Rosolowsky et al.~2003). 
As we discuss in the next section, our clouds are very far away from being
gravitationally bound. Velocity gradients have
also been observed in HI clouds associated
with circumstellar HI in the direction of evolved stars \citep{gerard06},
and have been explained in terms of expansion in the HI shell due to stellar mass loss.
%Similarly, a velocity gradient of $\sim$0.6 km~s$^{-1}$~arcmin$^{-1}$
%has been observed in a compact HI cloud by Dedes et al.~(2008). 
%In this case, the simplest explanation for the observed velocity gradient
%is if we are dealing with 
In this case, the observed clouds (or filaments) are likely to be inclined along the
line-of-sight (LOS) and the observed velocity gradient is most likely underestimated.
If a cloud/filament is inclined at an angle $\phi$ with respect to our LOS, 
and $\phi=0^\circ$ then no velocity gradient would be observed as all parts of 
the filament would lie along the LOS. Similarly, if $\phi=90^\circ$, 
velocities would be perpendicular to the LOS 
and hence no gradients observed. For any other value of $\phi$, the observed velocity
gradient will on average be less than the intrinsic velocity gradient.
This, together with the beam smearing effect, suggests that 
our measured average velocity gradient of $\sim0.5-1.0$
\kms~arcmin$^{-1}$ is likely to be underestimated and a more reasonable value
may be  $\sim2$ \kms~arcmin$^{-1}$.

For comparison, the velocity gradient seen in low mass dwarf galaxies
is typically greater than 2 km~s$^{-1}$~arcmin$^{-1}$ \citep{begum08,begum06}.
High resolution interferometric follow-up observations of compact clouds are
clearly needed, and are already underway, to provide detailed maps of
the clouds velocity fields.

\section{Derived cloud properties}
\label{sec:props}

\begin{figure*}
\epsscale{2.}
\plotone{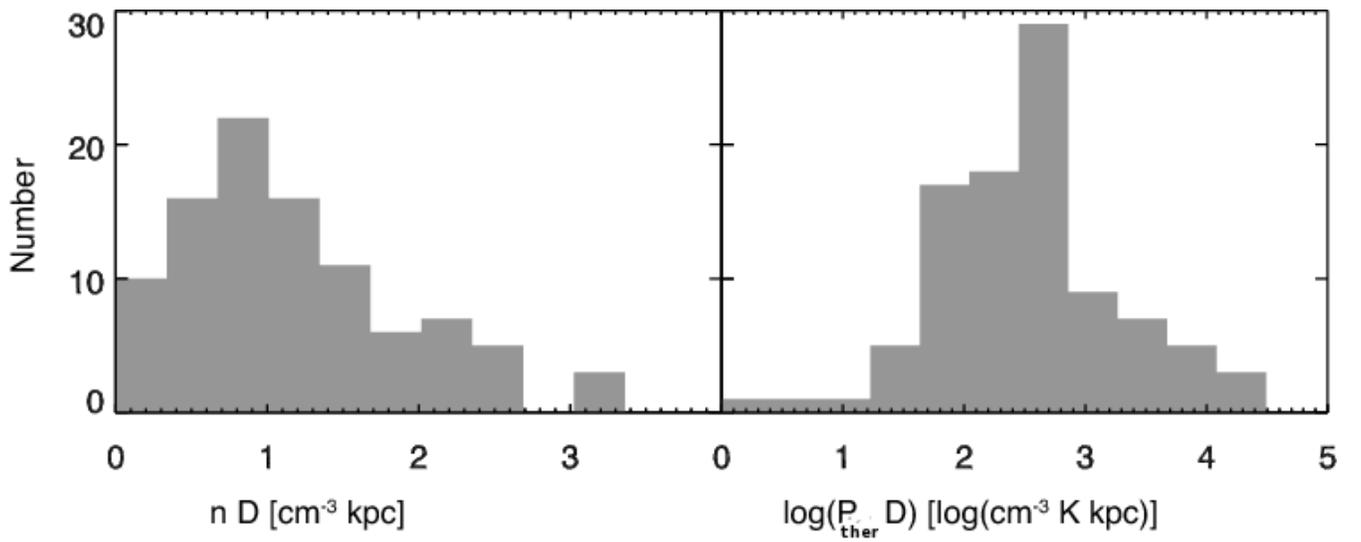}
\caption{\label{f:density} Histograms showing the HI volume density
and log(thermal pressure) distributions for the compact clouds.
}
\end{figure*}

We can use the measured cloud properties to derive the HI volume
density, the thermal pressure, and
the HI mass for the compact cloud population.
Assuming compact clouds to be spherical, the volume 
density is $n \propto {\rm{N_{HI}}}/({\rm {D}} \times \theta)$ and the
thermal pressure is $P_{\rm ther} \propto nT_{K, max}$,
with ${\rm{D}}$ being the unknown cloud distance.
We note that, as $T_{K, max}$ is derived from the observed FWHM, which
can have a contribution from  turbulent motions, the estimated
$P_{\rm ther}$ represents only an upper limit on thermal pressure. Also for
unresolved clouds the measured angular size $\theta$ is only a lower
limit on the cloud size. %Another free parameter in the derived cloud
%properties is the ratio of the
%thickness of the cloud along the line-of-sight to its width in the plane
%of the sky. We are assuming this ratio to be 1 i.e. the clouds are
%spherical. However
%if the clouds are thicker in the line-of-sight dimension,
%Also, most of the  clouds in our sample are no more %than a few
%beam-widths in size, so there is a significant  uncertainty in %their
%size estimates.  
Both $n$ and $P_{\rm ther}$ depend on the cloud distance D.
As cloud distances are unknown, we show in Figure~\ref{f:density}
quantities ($n~{\rm{D}}$) and log($P_{\rm ther}~{\rm{D}}$). To obtain true $n$ and
$P_{\rm ther}$ values, the derived quantities should be divided by the
distance (measured in kpc).
At a distance of 1 kpc, for example, $n\sim1$ cm$^{-3}$ and $P_{\rm ther}
\sim 300$ K~cm$^{-3}$.
We explore this concept further in later sections.

\begin{figure*}
\epsscale{2.0}
\plotone{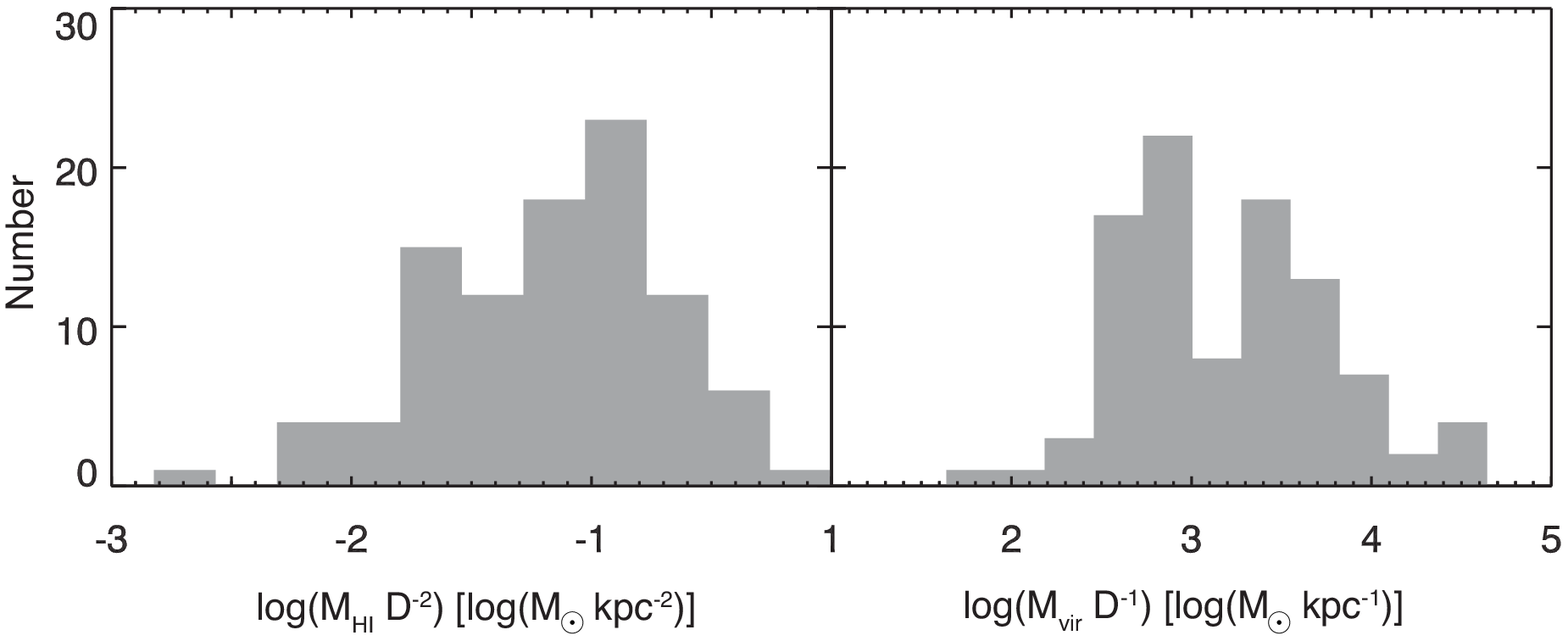}
\caption{\label{f:Mvir} Histograms showing the distributions of HI
mass and virial mass for the compact clouds.
}
\end{figure*}

We can now test the hypothesis that clouds are gravitationally bound.
The virial mass, M$_{\rm vir}$, is given as M$_{\rm vir}=190\Delta V^2 r$,
where $r$ is the cloud size in pc, $\Delta V$ is the velocity  FWHM in
\kms~and M$_{\rm vir}$ is
in units of solar mass. Figure~\ref{f:Mvir} (right) shows the
histogram of M$_{\rm vir} \times {\rm{D}}^{-1}$ for the whole cloud sample. The
median M$_{\rm vir} \times {\rm{D}}^{-1}$ for the sample is 2 $\times 10^3$
M$_{\odot}$ kpc$^{-1}$,
whereas the median HI mass for the sample is $\sim$0.1 M$_{\odot}$
kpc$^{-2}$ (see Figure~\ref{f:Mvir}, left). This implies that
${\rm{\frac{M_{vir}}{M_{HI}}}} \sim \frac{2 \times 10^4}{\rm D(kpc)}$. 
Hence, for D $\sim$  a few kpc, 
gravity is totally negligible, unless large amount of dark matter
is invoked to stabilize the clouds.

\section{Comparison with other surveys}

\subsection{HVC and IVCs}

In order to find an association, if any, between  the compact clouds
and well-known HVCs and IVCs, the compact cloud sample was
cross-correlated with the catalog of HVCs and IVCs
by Wakker (2001).  We find that a subset of clouds viz.~clouds No 48
to 57 with $l\sim 75^\circ$
and $b\sim -20^\circ$, lie close to a region having a large number of
scattered, faint IVCs.
This field is a part  of Complex GP HVC (Figure 18 in Wakker 2001),
having a distance limit of
0.8 kpc $< D < 4.3$ kpc. However, there is a $\sim30$ \kms~difference
between the lowest velocity found for the GP complex and the first
cloud we detect in this region.
Hence an association of these compact clouds with the Complex GP HVC is not very likely.

 Similarly, several  compact clouds around $l\sim 155^\circ$ and
$b\sim-30^\circ$ (clouds
No 13 to 21) lie close to  the Cohen Stream and HVC WW507 (Figure 9 in
Wakker 2001). The limit
on the distance to the Cohen Stream and WW507 is $>$ 0.3 kpc. Again,
as in the case of Complex GP,
there is a $\sim$ 25.0 \kms~difference between the lowest velocity gas
associated with this complex and our clouds detected in this region.
We therefore conclude that an association between compact clouds and
cataloged HVC and IVC
complexes is unlikely.

\subsection{Comparison with NaI and CaII absorption lines}

The compact cloud sample was cross-correlated with the recent
catalogue of NaI and CaII
absorption measurements towards 1857 early-type stars located within
800 pc of the Sun (Welsh et al. 2010). NaI and CaII absorption
relatively close to our clouds would allow us to place a constraint on
the cloud distance.
Only one possible correlation was found within our search radius of
$\sim$7 arcmin. In this case, cloud No 46 is located at an angular
separation of only $\sim$1.7 arcmin from
HD 151525 (at a distance of 141 pc). However, the V$_{\rm LSR}$ of the
NaI/CaII absorption is significantly different from
that of the cloud HI emission, suggesting that the cloud is likely to
be at D$\geq$141 pc.

\subsection{Variable stars}

Braun \& Kanekar (2005) and Dedes et al. (2008) suggested possible
shell-like structures associated with several small HI clouds.
As this points to a potential mass-loss cloud origin,
we correlated our cloud catalog with the catalogue
of variable stars (Downes et al.
2006\footnote{http://archive.stsci.edu/prepds/cvcat/}).
The catalogue includes novas, nova-like variables, Mira Variables,
interacting binary dwarfs etc.
For one of the compact clouds (Cloud No 31), a ``nova-like" variable
star (Com) was found
within 10 arcmin of the clouds, and for 12 clouds (i.e.~12.5\% of the
sample of clouds), there is a
variable star within 1 degree. If, on the other hand, we randomly
select 96 lines of sight, we find that
$\sim$7.0\% of those  lines-of-sight have at least one variable star
within a degree. This suggests that
a potential correlation between variable stars and compact clouds may be more than
a chance alignment.

\subsection{Optical/UV/IR extragalactic surveys}

For each detected compact cloud, the NASA Extragalactic Database (NED)
was searched at the location of the cloud center for possible
optical/UV/IR counterparts. The search was performed within 3.5 arcmin
of the peak of the cloud HI distribution. No diffuse
optical/UV emission was detected at cloud positions in the DSS, SDSS
or GALEX imaging surveys. For 39 compact clouds (i.e.
40\% of the sample of clouds), an extended IR source was found within
a search radius of  3.5 arcmin in the 2MASS extended source catalog.
Nothing is known about the nature of these IR sources; no optical
redshifts are known for any of the sources, hence their association
with the HI clouds is yet to be confirmed. However, this potential correlation
with the IR sources is more likely to be a chance alignment, as for
96 randomly selected lines of sight, $\sim$43 \% have at least
one 2MASS extended source within 3.5 arcmin.

\section{Distance constraints: Cloud Distribution Models}
\label{sec:cdm}

In this section we compare the distributions of compact clouds in Galactic
latitude ($b$), Galactic longitude ($l$), and radial velocity
(${\rm{V_{LSR}}}$) to models of cloud populations in order to constrain
physical characteristics of the population, such as distances and
intrinsic velocities.

\subsection{Galactic Latitude Distribution}
\label{gld}

By examining the distribution in Galactic latitude of the detected clouds,
we can determine whether they are linked to the Galactic disk: a
distribution that is oblate in the plane of the Galaxy would indicate
a relationship to the Galactic disk. The method is to construct a
simplified model of the distribution of these clouds and then compare
it with our observed distribution using the Kolmogorov-Smirnoff (K-S)
D-statistic. The D-statistic is simply the largest difference between
the cumulative distribution functions (CDFs) of the data and model. In
this way we can fit a parameter in our simplified model to the data.

We choose a model that captures the relative scale distance at which
these objects can be
seen in the plane of the Galaxy to the scale height of a population
above the disk.
We call this ratio of scales $R$. If the clouds are independent of the
orientation of the
Galactic plane, the K-S D-statistic should be at a minimum for a
spherical distribution,
$R=1$. If they do follow the plane, and can be seen out past their
Galactic scale height,
$R > 1$ should have a significantly smaller D-statistic. Our model
simply assumes that
clouds are equally detectable along the surface defined by
\begin{equation}\label{obl}
L^2 = x^2 + y^2 +  \left(Rz\right)^2
\end{equation}
where $L$ is an arbitrary distance, $x, y,$ and $z$ are distances from
the Sun, with $x$
toward the Galactic center and $z$ pointing towards the north Galactic
cap. We also assume
the probability of detecting a cloud at great distance drops to zero,
and that the functional
form by which it drops is the same along each axis. We then compute
the distribution of
clouds we expect to observe from this model as a function of $R$,
given the footprint of
the GALFA-HI data set, and compute the K-S D-statistic.

\begin{figure*}
%\epsscale{0.6}
\begin{center}
\includegraphics[scale=0.6,angle=0]{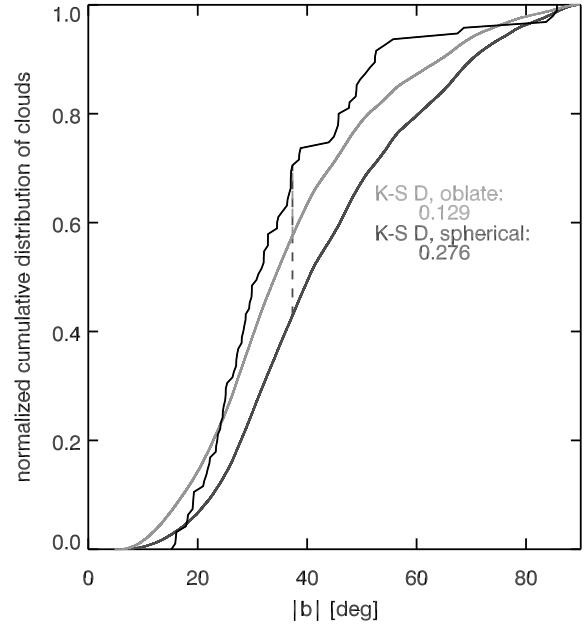}
\caption{\label{f:KSD}The normalized cumulative distribution of sample
clouds plotted as a function of Galactic latitude. The same quantity
is plotted for the two models with Galactic clouds having spherical
and oblate distributions. The dashed line shows where the D-statistic
is computed i.e where the difference between the model and the
observed cumulative distribution function is the largest.
}
\end{center}
\end{figure*}

As shown in Fig.~\ref{f:KSD}, we find that a spherical distribution is
a much poorer fit to the data ($D_{R=1} = 0.276$)
than an oblate distribution ($D_{R=10} = 0.129$), but that we cannot
distinguish between
distributions with $R \ge 10$. This is because we have very little
coverage in the Galactic
plane, where a very oblate distribution would generate a very large
number of clouds. Therefore,  we see evidence that significantly more
clouds are located toward the plane of the disk. This suggests an
oblate cloud distribution and implies that the bulk of the clouds are
related to the disk.
%With the future addition of the low-latitude regions we will be able to constrain this better.

As mentioned in Section~\ref{s:observations}, the Arecibo telescope does not allow an 
unbiased access to all Galactic latitudes. Currently, most of our coverage is at 
intermediate Galactic latitudes, $|b| \sim 20^\circ-50^\circ$ . We will add in the future data from lower 
latitudes: $b$=[-10:10] from I-GALFA (Inner-Galaxy ALFA  Low-Latitude H I Survey). This 
will allow us to constrain better the latitude  modeling and/or any potential variation 
in the cloud properties with Galactic latitude.

\subsection{Galactic Velocity Distribution}
\label{ssec:gvd}

We extend the analysis of the previous section by examining the
distribution of velocities.
In theory, it is possible to use the distribution of velocities to
determine whether the clouds
are corotating with the disk or whether they are independent of disk
rotation. This velocity distribution, interesting in its own right,
can also help shed light on the typical distances to the clouds.
In addition, modeling of the velocity distribution depends less on 
our incomplete survey coverage (Section~\ref{s:observations}) 
and offers a wider range of data points for the comparison of various distributions.

\subsubsection{Galactic Model}

In this model we generate a large number of clouds which populate
($l$, $b$, $v$) space, and have velocities related to the bulk
rotation of the Galactic disk.
To do this we assume the clouds populate an oblate distribution with
$R = 10$ and some
variable distance scale $L$ (see Equation \ref{obl}). We assume the
clouds follow cylindrical
Galactic rotation with a flat rotation curve at $\Theta_0 = 220$
kms$^{-1}$. In addition to
their rotation velocity, we impart to them a variable random radial
velocity, $\sigma$. We
constrain the clouds to lie in our observational footprint and to meet
the simple detection
criterion that they do not lie in regions with background brightness
temperature in HI above 4 K,
which is consistent with our completeness test (Section \ref{ssec:complete}).
We compute this background brightness using
the 36$^\prime$ resolution LAB survey (\cite{lab}).

\begin{figure*}
%\epsscale{0.3}
\begin{center}
\includegraphics[scale=0.6,angle=0]{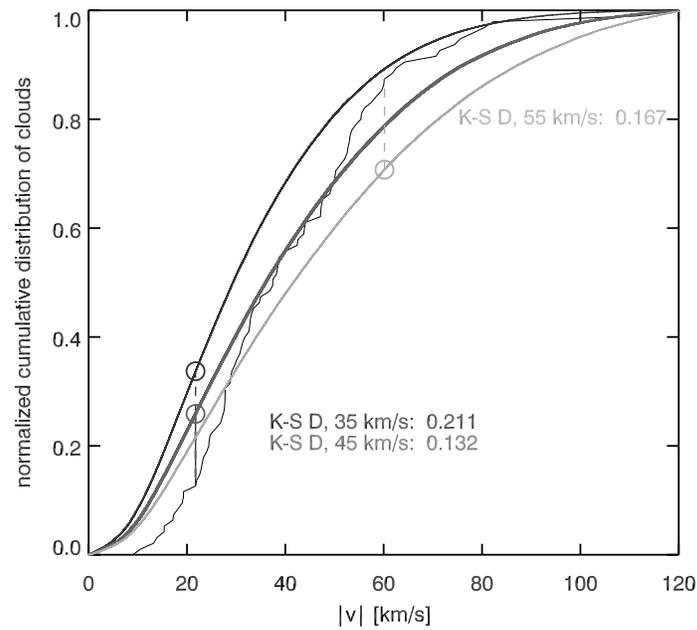}
\caption{\label{f:KSD1}The normalized cumulative distribution of
sample clouds plotted as a function of Galactic LSR velocity. Smooth
lines show modeled cumulative distributions for a varying random
velocity component. The dashed line
shows where the D-statistic is computed i.e where the difference
between the model and the observed cumulative distribution function is
the largest.
}
\end{center}
\end{figure*}

We find that comparing our model to the data does not constrain the
distance to the
clouds under the assumption of corotation. %Most of our clouds lie along
% $-40^\prime \leq b \leq -30^\prime$, %Since the clouds typically lie at
%$|b| > 30^\circ$, it is unreasonable to expect them to be in
%corotation past a distance
%of a few kpc (e.g. \cite{levine08}, \cite{collins02}). At these
%distances, the effect of
%corotation is rather slight, of order 10 kms$^{-1}$, which is
%completely swamped
%by the high random velocities of the observed clouds. It is therefore
%impossible to
%say anything about the distances to these clouds using the velocity
%information.
We can, however, fit to the random velocity component, $\sigma$,
independent of $L$ and
corotation. Using the K-S D statistic, as in Section \ref{gld}, we find
that a random radial
velocity of 45 \kms is near the best fit to the data, with a variation
from this value of
10 \kms~being noticeably worse (see Fig.~\ref{f:KSD1}). Assuming there
is no preferential
direction for this random velocity, we can multiply by $\sqrt{3}$ to
find a true 3-space
random velocity of $\sigma_3 \sim 80$ \kms.

\subsubsection{Circumgalactic Model}
\label{ssec:gsr}

\begin{figure*}
%\epsscale{1.0}
\begin{center}
\includegraphics[scale=1.0,angle=0]{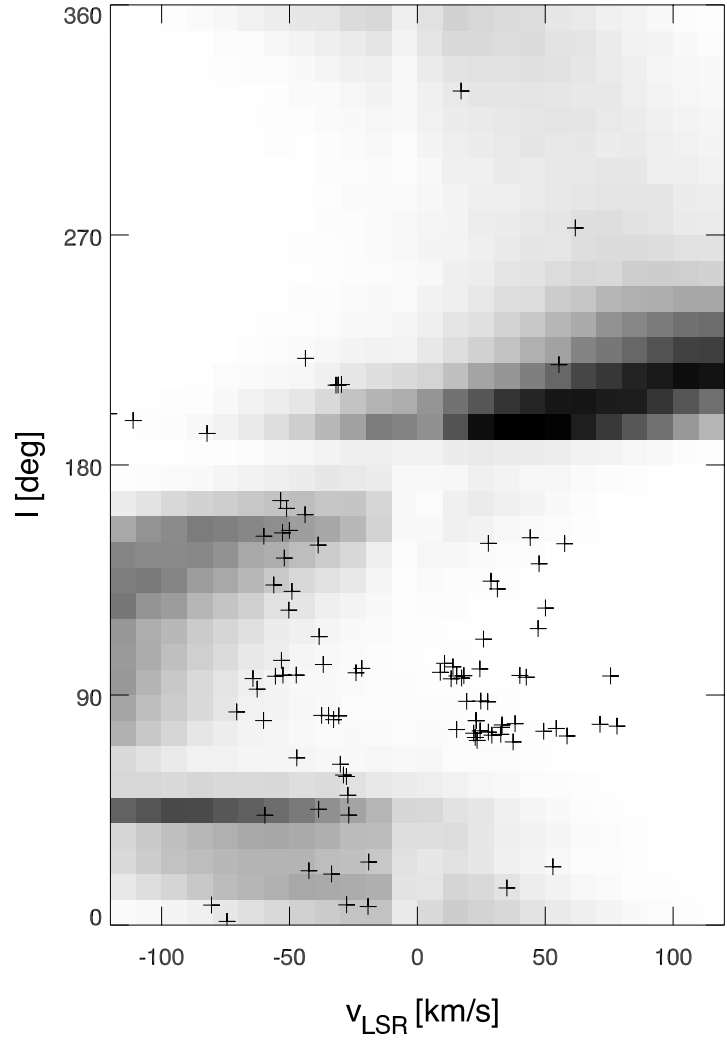}
\caption{\label{f:gsr} Galactic longitude as a function of V$_{\rm LSR}$.
The crosses show the sample compact clouds, whereas the greyscale
represents the circumgalactic model of the compact clouds (see text
for more details).
}
\end{center}
\end{figure*}

We can also generate a model with the basic assumption that the clouds
are beyond, but
around, the Galaxy, and therefore have $v_{GSR}$ velocities near to
zero (i.e.~they are in the rest
frame of the Milky Way's barycenter). As in the Galactic model, we
include the selection
effects of our data set when producing the model. We include a
$\sigma$ of 45 kms$^{-1}$
as in the Galactic model. We find that the circumgalactic model is
qualitatively a very
poor fit to the data (see Figure \ref{f:gsr}) -- many observed clouds
lie in areas where the
model predicts almost no clouds (e.g. ${\rm{V_{LSR}}} \simeq$ 30 \kms, $l
\simeq$ 90 degrees),
and the model predicts many clouds where none are observed (e.g.
${\rm{V_{LSR}}} \simeq$ $-90$ \kms,
$l \simeq 150$ \kms). We note that by increasing $\sigma$ we could
modify the model to predict more clouds in areas where there are
discrepancies, but this would have the effect of also producing more
model clouds out to extreme velocities, where clouds are not observed.
Based on this model we conclude that the observed clouds must be
related to the Galactic disk. This conclusion is consistent with, but
entirely independent of, our conclusion in Section \ref{gld}.

\section{Distance constraints: other methods}
\label{sec:tangent}

The distances to some of the cataloged clouds in the I$^{st}$ and
IV$^{th}$ Galactic quadrant can be measured using the tangent-point
method.
The LSR velocity at tangent points (${\rm{V_t}}$), positions where the line of sight
is perpendicular to a circle of constant Galactocentric radius, was
measured from $^{12}$CO observations by Clemens (1985).
 We find that 17 clouds in the I$^{st}$ Galactic quadrant have
${\rm{V_{LSR}-V_t }}<30$ \kms. If we assume that all these clouds are at the
tangent point but their velocities have been perturbed from $V_t$ by
random motions \citep{stil06b,ford08},
we can then estimate the clouds tangent point distance: d$_t=R_\odot
\cos l / \cos b$ (with R$_\odot=8.5$ kpc being the Galactocentric
radius of the solar circle). Under the further assumption that
Galactic rotation is constant with distance from the plane,
we find that d$_t$ ranges from 0.4  to 3 kpc, with a median distance
being 2.2 kpc. This would result in a median height from the Galactic
Plane h$_z=d_t \times \sin b \sim$1 kpc.
%The derived tangent point distances are consistent with distance
%constraints in Sec.\ref{sec:cdm}.

Another distance constraint comes from the method applied by
Stanimirovic et al.~(2006).
Compact clouds are cold with properties similar to what is found for
typical Galactic CNM clouds, or they show evidence for the existence
of cold cores surrounded by warm envelopes.
To explain the coexistence of the CNM and WNM, heating and cooling
conditions need to
be considered locally (Wolfire et al. 2003), and a well-defined range
of thermal pressures is needed. For example, at $R_g=8.5$ kpc,
$P_{min}\sim2000$ and $P_{max}\sim5000$ K cm$^{-3}$.
We can compare this pressure requirement with our pressure histogram
in Figure 14.
For a thermal pressure typical for the solar neighborhood of 3000 K
cm$^{-3}$, a distance of $\sim100$ pc is required. Wolfire at al.
(2003) showed that the lowest thermal
pressure allowed for the CNM and WNM to coexist is $\sim300$ K
cm$^{-3}$. This would
place an upper limit on the cloud distance of $\sim1-3$ kpc.
This is in agreement with the tangent point method and suggests a
possible distance range of 0.1 to 3 kpc.

The cloud ${\rm{V_{LSR}}}$ distribution in Figure 5, roughly symmetric with a
similar number of clouds having either negative or positive ${\rm{V_{LSR}}}$, 
places another constraint on the cloud distance. As most clouds in the sample
have $|b|>30$ degrees, a cloud distance $>3$ kpc would imply a height above the
Galactic plane that is $>1$ kpc.  At such heights, Galactic rotation is lower
than in the midplane of the Galaxy \citep{marinacci,kalberla09,levine08,collins02}.  
This results in the ${\rm{V_{LSR}}}$  distribution for the whole sample to be skewed
towards negative velocities. This effect  of a lag in clouds rotation velocity at 
high $|z|$ resulting in the LSR distribution getting skewed towards negative LSR 
velocities have been noted for the HVCs distribution \citep{peek09} and the 
Milky halo stars \citep{schneider06}. As we do not observe this
effect when considering the whole sample of compact clouds suggests that 
a majority of clouds are likely to be at a distance $\leq3$ kpc.

\section{Discussion}
\label{s:discussion}

\begin{table*}
\caption{Comparison of GALFA-HI clouds with the compact clouds from other Galactic surveys/studies}
\centering
\label{tab:compare}
\begin{tabular}{c c c c c c}
\hline
\hline
    \multicolumn{1}{c}{Parameters} &
    \multicolumn{1}{c}{GALFA-HI} &
    \multicolumn{1}{c}{GASS} &
    \multicolumn{1}{c}{VGPS} &
    \multicolumn{1}{c}{Lockman clouds} &
    \multicolumn{1}{c}{VLA/WSRT } \\
        & (This work) & \citep{ford08} & \citep{stil06b} &\citep{lockman02}  & \citep{bekhti09}  \\
\hline
N$_{\rm HI} (10^{19})$&0.5 & 1.4 &  18.0 &2.0  &0.9 \\
T$^{Pk}_B$ (K)& 0.75 & 0.6 & 12.3 &1.0  &1.6 \\
$\Delta$V (kms$^{-1}$) &4.2 & 12.8 & 5.6 & 12.2 & 4.1 \\
Size (arcmin)& 5.0 &29.0 &  3.4 & 22.3&2.3  \\
%${\rm \theta_{FWHM}}$& 3.5 &15.0 &  1.4 & 9.2&1.6  \\
%V$_{LSR}$ range (\kms)& $[-120:120]$ & $[-180:-70]$ & [60:160] & [103:148] & $[-120:-10]$   \\
\hline\end{tabular}
\end{table*}

Based on the modeling of spatial and velocity distributions
of the whole compact cloud population,
we have shown that the bulk of the compact clouds are related to the
Galactic disk.
The tangent point method and the consideration of the conditions required
for the CNM/WNM coexistence in the ISM suggest that cloud distances
are likely to be in the range  of 0.1 to a few kpc. However, this
still does not exclude the possibility for a few clouds to be located
at very large distances. As possible scenarios for the cloud origin
depend on distance,
we discuss several possibilities in the following sub-sections.

\subsection{Galactic compact clouds at sub-pc scales}

For a distance $\leq1$ kpc, with a typical size of $\leq1$ pc,
we are dealing with nearby Galactic clouds.
In this case,  a significant fraction of clouds would have
$n \geq$ 1 cm$^{-3}$ and a total pressure $\sim$ 1000 K cm$^{-3}$.
The cloud HI mass (shown in Figure~\ref{f:Mvir}, left) would be
$<10^{-1}$ M$_{\odot}$.
However, at a distance as low as 100 pc these properties become
slightly more extreme:
$n \sim$ 10 cm$^{-3}$, $P_{\rm ther} \sim$ 3000 K cm$^{-3}$,
${\rm{M_{HI}}}\sim10^{-3}$ M$_{\odot}$, and
clouds would have a linear size of $\leq30,000$ AU.
At this distance and assuming a velocity gradient of 2
\kms~arcmin$^{-1}$, the length doubling timescale for clouds/filaments
would be very short,  $\sim1.5\times10^4$ yr. Such short lifetimes
suggest transient and recently produced objects.

In any case, the properties (size, temperature, HI column density, HI
mass) of such sub-pc clouds may be similar to those of the low-column
density absorbers \citep{bk05,snez05}.
This possible connection is also supported by the WSRT HI emission
observations by Braun \& Kanekar (2005),
who found several compact clouds in the proximity of one of the low
column density absorbers
seen in HI absorption. In this scenario, the compact clouds may
represent large and low column-density examples of the population of
CNM clouds on sub-pc scales. The most extreme CNM clouds, so called
``tiny scale atomic structure'' (TSAS),
reach scales of a few AU, appear over-dense and over-pressured
relative to the traditional CNM, and were once  postulated
to comprise up to 15\% of the CNM (Frail et al.~1994;
Heiles 1997).
The obvious questions are, how do such small and isolated clouds form
and survive in the ISM, and what role do they play within the general ISM?

One possible scenario that can explain the formation and maintenance of
compact cold clouds in the ISM is interstellar turbulence.
Recently numerical simulations of the ISM have started to describe cold
and warm atomic gas with a numerical resolution and dynamic range
approaching realistic physical scales e.g. \citet{Semadeni06,avillez05,Audit05}.
The numerical simulations by Audit \& Hennebelle (2005) show that a
collision of incoming  turbulent flows can initiate condensation of the WNM
into cold neutral clouds. A collision of incoming WNM streams creates a
thermally unstable region of higher density and pressure but lower temperature,
which further fragments into small  cool structures. The abundance of
cold structures,
as well as their properties, depends heavily on the properties of the underlying
turbulent flows. For example, Audit \& Hennebelle (2005)
find that a significant fraction of the CNM structures formed in the case of
very turbulent flows have  thermal pressures of  $\leq 10^4$ K cm$^{-3}$,
temperatures $\geq$100 K and volume density n$\leq 100$ cm$^{-3}$. These CNM
structures are thermally stable, long lived ($\sim$ Myr) and in the case of
stronger turbulence
they are more rounded, similar to the compact clouds in our sample.

In a similar approach, Koyama \& Inutsuka (2002) showed that a thermally unstable
shock-compressed layer, formed from the WNM compression by supernova explosions, can 
also fragment into small, cold,  turbulent condensations as a result of the thermal instability. 
As the ISM is frequently 
compressed by supernova explosions, fragmentation of the shock-compressed medium 
can also be a potential mechanism for the formation of compact clouds.

Another possibility may be provided by stellar outflows.
HI emission has been detected in the circumstellar shells of a variety of
evolved stars, viz.~asymptotic giant branch stars, oxygen-rich and
carbon-rich stars,
semi-regular and Mira variables, and planetary nebulae \citep{gerard06}.
For example, Matthews et al.~(2008) found an extended tail of HI associated with the
AGB star Mira and argued that HI tails are likely to be  a common
feature of evolved stars undergoing mass loss.
Furthermore, HI has also been detected in emission/absorption
toward several planetary nebulae (Gussie \& Taylor 1995, Rodriguez et al.~2002).
Our cross-correlation of the compact cloud catalog with the
catalogue of variable stars by Downes et al.~(2006)
suggests that a subset of clouds has at least one variable star within
a radius of 1 degree.
%\footnote{http://archive.stsci.edu/prepds/cvcat/}).

The association
of HI clouds with variable stars at angular distances of $\sim1-2$
degrees has been noticed before.
Matthews et al. (2008) found HI emission as far as $\sim 1.5^\circ$
away from a Mira variable.
Gerard \& Le Bertre (2006) also reported evidence that HI emission
associated with circumstellar envelopes may be offset from the
position of the central star. Assuming that the detected HI clouds are
the circumstellar HI associated with variable stars, for a typical
distance
of 100 pc and an expansion velocity of 5 \kms~seen for circumstellar
HI \citep{gerard06},
a degree separation of the HI clouds from the variable star
corresponds to an HI mass of $\sim 2 \times 10^{-3}$ M$_\odot$, an HI
diameter of 1.7 pc, and a characteristic time-scale
of 0.34 Myr. All three parameters agree well with the ones found in
the HI survey of circumstellar
envelopes around evolved stars \citep{gerard06}. Hence it is likely
that some of the compact HI clouds could be related to the outflows
from these stars.

Once formed by turbulence and/or stellar outflows and injected into the
surrounding  medium, these isolated compact clouds of cold, low column
density HI will be immersed in the warm/hot ambient gas. The obvious
question that arises is
how are such compact clouds able to survive without
being quickly evaporated. From Eq.~(47) in \citet{mckee}, the mass-loss of a compact HI
cloud embedded in a hot plasma ($T\sim 10^6$ K) is
$\sim 7 \times 10^{-2}$ M$_\odot$ Myr$^{-1}$, whereas if the clouds
are embedded in
large warm envelopes ($T\sim 7000$ K), the evaporation mass-loss is
ten times smaller.  This would imply that the compact clouds may be
evaporating on a timescale of $\sim$ Myr, due
to a combination of conductive heat transfer and/or Kelvin-Helmholtz
instabilities
from the surrounding warm/hot medium \citep{snez05,dedes10}.

\subsection{Compact clouds in the disk-halo interface region}

If at a distance of a few kpc, the majority of compact clouds
would have  $P_{\rm ther} \leq$ 100 K cm$^{-3}$, $n\sim$ 1.0 cm$^{-3}$,
a height of a few kpc above the disk, and a corresponding HI mass of
$\sim 0.01 - 0.3$ M$_\odot$.
In this scenario, observed cloud properties are similar to
numerous HI clouds found in the Galactic disk-halo interface region.

Recent HI observations have shown that the disk-halo
interface of our Galaxy is not smooth, but populated with 
discrete HI clouds.
Observations with the Green Bank telescope (GBT) found discrete HI clouds in
the halo of the inner Galaxy (Lockman 2002). Although these clouds,
often referred to as ``Lockman's clouds'', are located $\sim$ 900 pc
below the plane, they  appear to be dynamically related to the disk.
Further, \cite{stil06b} discovered many clouds at
forbidden velocities in the inner Galaxy using data from the
VLA Galactic Plane Survey (VGPS). Despite their forbidden velocities, those
clouds follow the Galactic emission. Similarly, Ford et al.~(2008)
detected over 400 HI clouds in the lower halo of the inner Galaxy using the
Galactic All-Sky Survey (GASS), with properties similar to the clouds
discovered by Lockman (2002). \cite{snez06}, using the GALFA-HI precursor
observations, found numerous clouds in the disk-halo interface of the
outer Galaxy, indicating that this phenomenon is 
not restricted to the inner Galaxy. Finally, using the
interferometric data from the Westerbork Synthesis Radio Telescope
(WSRT) and the Very Large Array (VLA), \cite{bekhti09} studied
the HI emission at low, intermediate and high radial velocities along
four random lines of sight toward quasars which had  previous detections of
weak CaII and NaI absorption features.
They found several compact and cold clumps of neutral gas at
velocities similar to those of the optical absorption.

Table~\ref{tab:compare} shows a comparison of the compact clouds from
this work with the clouds discovered in the recent Galactic surveys. 
Based on their physical parameters,  N${\rm{_{HI}}}$, T$^{Pk}_B$ and $\Delta$V, 
compact clouds from the GALFA-HI survey are the most similar to the low-column
density HI clumps seen in the HI interferometric observations along
the sight-lines of CaII and NaI absorbers (Bekhti et al. 2009).
However, most differences between various studies stem from the difference
in angular resolution and sensitivity of various surveys. 
For example, higher values of N${\rm{_{HI}}}$ and T$^{Pk}_B$ for the \cite{bekhti09} 
clouds compared to our sample are mainly due to the difference in
angular resolution of the Arecibo (3.5 arcmin) and WSRT/VLA observations ($\sim$1.6 arcmin).
Larger size and velocity line-width of the GASS and GBT clouds samples
are largely caused by large telescope beams (15 arcmin and 9 arcmin for the Parkes  telescope and
the GBT, respectively) and the blending of smaller clouds. 
The HI clouds in the disk-halo interface region are generally thought
to originate from the  condensation of hot gas expelled from the disk
by superbubbles \citep{hb90}.
%The existence of discrete small-scale HI clouds in the interface region
%have been predicted by various theoretical models such as Galactic
%fountain and Chimney models, as well as in the numerical simulations
%(de Avillez 2000). 
Hence, these compact clouds in the disk-halo interface
could play an important role in studying the mass circulation
between the Milky Way disk and the halo.

An alternative mechanism that may be able to explain the existence of compact
clouds in the disk-halo interface region was recently proposed by 
Heitsch \& Putman (2009), and involves re-cooling of inflowing gas. 
Simulations show that as warm/ionized clouds
approach the Galactic disk, they encounter denser and  denser
material, which leads to the compression of clouds and causes them to
slow  down and recool, forming intermediate and low-velocity clouds
close to the Galactic disk. The compact clouds in our sample may be
able to fit into this framework, thus representing a later stage of
the infall process for HVCs.
The observed compact clouds may be getting integrated and digested by
the disk, and hence provide a smooth buildup of fresh star formation
fuel for the Galactic disk. 
The  random radial velocity component of 45 \kms~ for the clouds found
in Section 7.2.1, corresponds
to T$_{k,max}\sim4.4 \times 10^4$ K, and may be interpreted as compact
clouds being condensed out of an extended warm/hot layer in the
disk-halo interface as traced by O~VI absorption lines \citep{savage}.
Future metallicity and dust-to-gas ratio
observations of compact clouds would be the crucial test of the cloud infall
hypothesis.

\subsection{Compact clouds at extreme distances}

As discussed in Section \ref{sec:props}, if the compact clouds
are self-gravitating
systems, they have to be extremely dark matter dominated in order to be stable,
having  more than $\sim$99\%  of the total mass as dark. The presence
of a large number of such low mass, dark matter halos have been
predicted by cosmological $\Lambda$CDM simulations
(Klypin et al.~1999, Moore et al.~1999). \cite{st02} showed that most
of the gas in low mass halos should be found in a thermally stable,
ionized/WNM phase, within which cold cores may be able to form.
However, such models require the clouds to be at large distances
($\geq$100 kpc). We note that  although the compact cloud sample as a
whole cannot be circumgalactic, from our statistical analysis in Section
\ref{ssec:gvd}, we cannot rule out
the possibility that some of the clouds could be at large distance,
and hence they could be associated with the dark matter mini halos in
the vicinity of the Milky Way, as predicted  by cosmological
$\Lambda$CDM simulations.
 Recently, Giovanelli et al.(2010) found a set of isolated, HI sources
in the ALFALFA survey and argued that these could be associated with
isolated mini-halos in the outskirts of the Local Group.
Again, metallicity and dust-to-gas ratio observations of these clouds 
should be able to address the  mini-halo possibility.

\section{Summary and Conclusions}

The Galactic Arecibo L-band Feed Array HI (GALFA-HI) survey
is successfully mapping the entire Arecibo sky at 21-cm.
The survey covers a velocity range of $-$700 to 700 kms$^{-1}$ (LSR)
at a velocity resolution of 0.18 kms$^{-1}$ and
an angular resolution of 3.5 arcmin. The unprecedented resolution and
sensitivity of the GALFA-HI survey resulted in the detection of
numerous isolated, compact HI clouds at low Galactic velocities,
which are distinctly separated from the disk HI emission. In the
limited area of $\sim$4600 deg$^2$ surveyed so far, we have detected 96 such
compact clouds. The detected clouds
are cold with median T$_{k,max}\sim300$ K. Moreover, these clouds are quite compact and
faint with median properties of 5 arcmin in angular size, 0.75 K in
peak brightness temperature, and N$\rm{_{HI}}$ of  $5\times10^{18}$
cm$^{-2}$. Most of the clouds deviate  from Galactic rotation at the
20-30 kms$^{-1}$ level, and a significant fraction shows evidence for a
multiphase medium and/or velocity gradients. No counterparts for these
clouds were found in other wavebands.

From  the modeling of spatial and velocity distribution of the whole
compact cloud population, we find that the bulk of the population is
related to the Galactic disk and their distances are likely to be in
the range of 0.1 to a few kpc. This is consistent with distance
estimates based on the thermal pressure requirements for the CNM and
the WNM coexistence, and the tangent point
method for a subset of clouds.

If the clouds are at a distance $\leq$1 kpc,  they would represent
low-column density examples of the population of CNM Galactic clouds
on sub-pc scales. Possible mechanisms for the formation of such 
clouds involve stellar outflows, and/or
condensation of the WNM by the collision of turbulent flows.  Once
formed, such cold, low  column density, compact HI clouds would
evaporate on timescales of $\sim$ Myr due to the combination of
conductive heat transfer and/or Kelvin-Helmholtz instabilities from
the surrounding warm/hot medium. Conversely, if the clouds are at
distances of a few kpc, they are similar to the
HI clumps observed in the Galactic disk-halo interface region. 
One potential mechanism that may be able to explain the existence of such cold 
clouds was recently proposed by \cite{heitsch09} and involves
the re-cooling of inflowing gas.
Finally, our modeling of cloud distributions cannot exclude the
possibility of sporadic compact clouds being at large distances. If 
being at large distances and self-gravitating, such clouds would 
be extremely dark matter dominated and may be related to the dark 
matter mini-halos predicted by cosmological simulations.

Our completeness check of the compact cloud catalog, showed that we are most likely 
missing about one quarter of the clouds due to our selection biases for regions where 
Galactic background is below 4 K. In addition, we cannot detect any clouds in regions 
where the Galactic background exceeds 4 K.  We speculate that faint, compact HI clouds 
are probably  widespread throughout the Galaxy, but we are able to detect them only in 
the regions with low Galactic emission.  In the future, by completing the GALFA-HI survey 
and applying automated methods for cloud detection we will collate a larger sample of 
compact clouds and extend our latitude coverage to $b\sim0$ degrees. 
This will allow us to better constrain cloud spatial distribution
through the latitude and velocity modeling discussed in the paper.
In addition,  we will investigate whether compact clouds are 
related to local events, such as stellar winds or
large-scale atomic flows, or are globally distributed
across the disk with notable kinematic properties.
To fully study  this cloud population we will have to wait for the next generation 
radio telescopes (Australian SKA Pathfinder, MeerKAT, or the Square Kilometer Array) which
will play an important role in providing a census of compact clouds by providing sensitive 
and high-resolution surveys, capable of filtering large-scale diffuse HI emission.

%the generation of additional catalogs with the ongoing incoming data is in progress.

\begin{acknowledgements}
We are grateful to the staff at the Arecibo observatory,
as well as the ALFALFA team (Giovanelli et al. 2005),
for their help in conducting the GALFA-HI observations. 
We thank Barry Welsh for his help with the NaI and CaII absorption lines.
A.B., S.S., M.E.P., C.H., E.J.K., and J.E.G.P. acknowledge
support from NSF grants AST-0707597, 0917810,
0707679, and 0709347. K.A.D. acknowledges funding from the European
Community's Seventh Framework Program under grant
agreement n$^{\rm o}$ PIIF-GA-2008-221289.
J.S.G. thanks the Graduate School of the University of 
Wisconsin-Madison for support of his research in this area.
We credit the use of the KARMA visualization software (Gooch 1996).
The Arecibo Observatory is part of the National
Astronomy and Ionosphere Center, which is operated by Cornell
University under a cooperative agreement with the National
Science Foundation.
\end{acknowledgements}

%\bibliographystyle{/rattler/sstanimi/TexStyles/apj}
%\bibliography{/rattler/sstanimi/Thesis/Thesisdir/bib/astro-mnemonic,/rattler/sstanimi/Thesis/Thesisdir/bib/myref}

\label{lastpage}
\end{document}